\documentclass{aa}

\usepackage{graphicx}
\usepackage{adjustbox}
\usepackage{rotating}
\usepackage{placeins}
\usepackage{float}
\usepackage{pdflscape}
\usepackage{hyperref}
\usepackage{txfonts}
\usepackage{multicol}
\usepackage{placeins}
\usepackage{sidecap}
\begin{document}

   \title{A song of interactions and mergers: The case of NGC~4709\thanks{This paper includes data gathered with the $6.5$ meter \textit{Magellan} Telescope located at Las Campanas Observatory, Chile, and obtained with MegaCam (\citealt{Megacam}). The data products are produced by the OIR Telescope Data Center, supported by the Smithsonian Astrophysical Observatory.}}

\titlerunning{A song of interactions and mergers}

   \author{
S. Federle\inst{1,2}\thanks{E-mail: sara.federle89@gmail.com}, M. G{\'o}mez\inst{1}, S. Mieske\inst{2}, F. Dux\inst{2,3},  M. Hilker\inst{4} \& I.~A. Yegorova\inst{2}}
\authorrunning{S. Federle et al.}

\institute{$^1$Instituto de Astrofisica, Depto. de Fisica y Astronomia, Facultad de Ciencias Exactas, Universidad Andrés Bello. Av. Fernandez-Concha 700, Las Condes, Santiago, Chile. \\
$^2$European Southern Observatory, Alonso de Cordova 3107, Vitacura, Santiago, Chile \\
$^3$Institute of Physics, Laboratory of Astrophysics, \'Ecole Polytechnique 
F\'ed\'erale de Lausanne (EPFL), Observatoire de Sauverny, 1290 Versoix, 
Switzerland \\
$^4$European Southern Observatory, Karl-Schwarzschild-Strasse 2, 85748 Garching bei M{\"u}nchen, Germany \\
}

   \date{}
 
\abstract  
{Globular clusters (GCs) are fundamental tools to unveil the interaction and merger history of their host galaxies.} {Our goal is to perform the photometric analysis of the globular cluster system (GCS) of the elliptical galaxy NGC~4709, which is the brightest galaxy of the Cen~45 spiral-rich galaxy group, and to highlight its interaction history with NGC~4696, the giant elliptical galaxy of the Cen~30 subcluster.}{We obtained deep {\em Magellan} $6.5$~m/MegaCam $(g', ~r',~i')$ photometry, with which we identified a sample of $556$ GC candidates around NGC~4709 that were analyzed in the context of the interaction history with the giant elliptical NGC~4696 and other galaxies of the Cen~45 group. After modeling and subtracting the galaxy light, we used criteria based on shape, colors and magnitude to select GC candidates. }{Our results point toward a complex interaction history that shaped the GCS of NGC~4709. Inside a galactocentric radius $r<5\times r_{\mathrm{eff}}$, two populations were found with mean colors of $(g'-i')_0=0.905\pm 0.009$~mag and $(g'-i')_0=1.170\pm 0.008$~mag. The azimuthal distribution of the GCs show peaks at the position angles ${\mathrm{PA}}_1=92^{\circ}$ and $\mathrm{PA}_2=293^{\circ}$, with $\mathrm{PA}_2$ coinciding with the direction toward NGC~4696, confirming that the interaction between these galaxies shaped the GCS of NGC~4709. From the GC luminosity function we derived a distance of $29.9\pm 2.1$~Mpc, which is much closer than the other galaxies of the Centaurus cluster, and a specific frequency of $S_N=3.7\pm 0.5$, in good agreement with previously estimated values. From the GCs density maps, we identified overdensities corresponding to the positions of five other galaxies of the Centaurus cluster, and we found a bridge of GCs between NGC~4709 and NGC~4696 with a distance of $d=34.69\pm 2.21$~Mpc, which is between that of the two galaxies.}{All of these findings point toward a complex GCS for NGC~4709, strongly influenced by the interaction with NGC~4696, and confirm previous findings that the galaxy's apparent distance is smaller than that of the main cluster galaxy NGC~4696 by $\sim 8.5$~Mpc and that it is smaller than the distance of the other Centaurus' galaxies, making it an outlier in Centaurus and suggesting a past first encounter with Cen~30. In future work, we will combine our data with E-MOSAICS simulations to unravel the trajectory of NGC~4709, from the past interaction to the future new rendezvous with NGC~4696. }


   \maketitle

\section{Introduction}
Globular clusters (GCs) are important tools to study the matter distribution and interaction history of their host galaxies. They are defined as compact, gravitationally bound systems of stars, which have a typical mean mass of $\sim 2\times 10^5$~M$_{\odot}$ (\citealt{Beasley20}). Even though some GCs in the region of the Galactic center (\citealt{Minniti21}) and disk (\citealt{Binney17}) are missing due to heavy obscuration by interstellar dust, the study of the globular cluster system (GCS) of the Milky Way can be used as a template for the analysis of extragalactic GCs. In particular, the GCS of the Milky Way contains $170$ GCs (\citealt{Vasiliev21}; \citealt{Ishchenko23}), with $\sim 300$ recently discovered candidates (e.g., \citealt{Garro21}; \citealt{Dias22}), thanks to surveys such as the VISTA Variables in the Via Láctea (VVV; \citealt{Minniti10}) and the VISTA Variables in the Via Láctea eXtended (VVVX; \citealt{Minniti18}). These GCs show a bimodal color distribution (e.g., \citealt{Renaud17}) and metallicity distribution (e.g., \citealt{Garro24}), with the metal-rich GCs being more concentrated toward the Galactic center, whereas the metal-poor GCs extend out to $145$~kpc in the Galactic halo (\citealt{Beasley20}). In the Milky Way, we can resolve GCs into their single stars out to $\sim 20$~Mpc with space-based missions such as the Hubble Space Telescope (HST), whereas with ground-based telescopes, as we go outside the Local Group, they appear as point-like sources. The first study of the GCS of M87 (\citealt{Racine68}) showed a large number of GC candidates, and more recent studies have placed it among the richest GCSs (\citealt{Oldham16}). Since then, GCSs have been studied in other galaxies, both in groups (e.g., \citealt{Urbano24}; \citealt{Obasi23}; \citealt{Taylor17}) and clusters (e.g., \citealt{Lomeli25}; \citealt{Janssens24}; \citealt{Federle}; \citealt{Puzia02}). Two scenarios seem to emerge: the presence of a single or of multiple populations of GCs highlighted from the uni-, bi-, or multimodality of the color distribution. For example, in the case of NGC~4262, a color bimodality was observed, and the properties of the GCS suggest that the galaxy might be transitioning into an elliptical \citep{Akhil24}. On the other hand, for the GCS of NGC~4365 a trimodal color distribution was identified with blue, "green", and red GC populations (\citealt{Puzia02}). The origin of the "green" GC population is still uncertain, but it could be the result of a merger with a gas-rich galaxy resulting in the formation of new GCs or it could have been stripped via gravitational interaction from the nearby S0 galaxy NGC~4342 \citep{Blom14}. Other examples are the GCS of NGC~4382, for which, besides two old GC populations, a young one with an age of $2.2\pm 0.9$~Gyr was observed (\citealt{Escudero22}), and NGC~1316 in the Fornax cluster where, besides the blue and red populations, an intermediate one with an age of $\sim 5$~Gyr and a very young population with an age of $\sim 1$~Gyr were observed (\citealt{Richtler14}; \citealt{Sesto16}). 

It has been shown that the relation between the number of GCs and the stellar mass of the host galaxy is nonlinear, with low-mass galaxies (especially dwarfs) and very massive galaxies being more prone to produce GCs than stars with respect to galaxies of intermediate masses (\citealt{Beasley20}). Moreover, the specific frequency of the GCSs depends on the galaxy's morphology, with the highest values found in giant and dwarf ellipticals, whereas the lowest values belong to late-type galaxies (e.g., \citealt{Harris91}; \citealt{Georgiev10}; \citealt{Alamo-Martinez13}; \citealt{Obasi23}). However, some exceptions arise in galaxy groups, where the loss of GCs through interactions results in early-type galaxies having a low specific frequency, such as in the case of NGC~5018 \citep{Lonare25}. It has been shown that the GCSs can extend from the inner regions of a galaxy to the outer haloes, with a halo population particularly numerous in early-type galaxies (\citealt{Reina-Campos22}), where they can reach galactocentric distances of $5-20~r_{\rm eff}$ (\citealt{Reina-Campos23}). This allows to study the kinematics and mass distribution of the host galaxies at large distances from the galactic center, where it is not possible to obtain 2D kinematic models of the stellar light, which for most galaxies is limited to $1~r_{\rm eff}$, and up to $2-4~r_{\rm eff}$ for some early-type galaxies (e.g., \citealt{Proctor09}; \citealt{Dolfi21}). Galaxies in clusters are more likely to undergo multiple major and minor mergers. Despite the fact that the velocity dispersion is large, the infall of galaxy groups provides a mechanism that promotes slow encounters and mergers within the cluster (\citealt{Mihos04}). Moreover, simulations of tidal interactions in galaxy clusters highlight the tendency of massive galaxies to be strongly clustered, so the probability of galactic collisions and mergers increases (\citealt{Gnedin03}). These merger events affect not only the morphologies of the galaxies but also of their GCSs. Mergers can account for debris such as tidal tails, which are large elongations on one side of a galaxy, that are particularly evident in the Antennae galaxies (e.g., \citealt{Lahén18}). Such features show the presence of dynamically linked groups of GCs that can give information on the interaction history of their host galaxies, as demonstrated in the cases of M~31 (\citealt{Veljanoski14}) and NGC~5128 (\citealt{Hughes23}). Tidal features were also identified photometrically, and are characterized by the presence of GCs, which can give more insights into the accretion history of the host galaxy, such as in the Dorado group (\citealt{Urbano24}). Merger events can also account for different populations of GCs, as was found in the low-mass early-type galaxy NGC~4150 (\citealt{Kaviraj12}). In fact, models have predicted that the red population of GCs would form in situ and would be more concentrated toward the bulge of the host galaxy, whereas the bluer GCs represent the accreted components and would reside in the halo (\citealt{Cote98}). This was confirmed by MOdelling Star cluster population Assembly In Cosmological Simulations within EAGLE (E-MOSAICS; \citealt{Pfeffer18}) simulations, where we can see that in situ and accreted GCs have comparable numbers in the inner parts of the halo, whereas GCs in the outskirts of the galaxies have a preferentially accreted origin (\citealt{Reina-Campos22}).

The Centaurus cluster is an ideal target to study such features. It is composed of the subcluster Cen~30, which represents the main component dominated by the giant elliptical NGC~4696, and Cen~45, a spiral-rich subcomponent dominated by NGC~4709 (\citealt{Lucey86}). The velocity distribution of cluster galaxies reveals the young, unrelaxed nature of the system, with Cen~45 being an infalling galaxy group (\citealt{Stein97}). Interactions between galaxies were confirmed in the central region of Centaurus by the presence of a filamentary structure connecting NGC~4696 to NGC~4696B, by a metallicity excess around NGC~4709 (\citealt{Walker13}), and by asymmetric temperature variations in the X-ray gas with the hottest regions coinciding with NGC~4709 (\citealt{Churazov99}), which is consistent with shock-heated gas due to the past interaction between Cen~45 and Cen~30 \citep{Veronica25}. Furthermore, NGC~4696 is characterized by a filamentary structure crossing its central regions and extending toward NGC~4696B, suggesting that it underwent some merging phenomena in the past (\citealt{Churazov99}) or that the material was acquired by ram-pressure stripping of NGC~4696B (\citealt{Walker13}). Interactions between galaxies in Centaurus are also reflected by the shape and properties of their GCSs. In \citet{Federle}, we studied the GCS of NGC~4696 and determined that it presents a bimodal color distribution, with the blue and red populations divided at a color $(g'-i')_0=0.905$~mag, and indications of the presence of an intermediate age GC population in the central regions of the galaxy. Moreover, the analysis of the azimuthal distribution of the GC candidates shows peaks coinciding with the directions of NGC~4696B and NGC~4709, confirming that the interactions between these galaxies shaped the GCS of NGC~4696. Finally, from the GC luminosity function, we found a distance of $d=38.36\pm 2.49$~Mpc for NGC~4696, which is consistent with the distance of $42.5\pm 3.2$~Mpc from \citet{mieske}.

In this paper, we expand our work done in \citet{Federle} to Field~1 of the Centaurus cluster (Fig.~\ref{fig:Centaurus}), with particular attention given to the characterization of the GCS of NGC~4709, whose properties are reported in Tab.~\ref{tab:galaxy}. The paper is organized as follows. In Sec.~\ref{sec:data} we describe the observation and data reduction, in Sec.~\ref{sec:selection} we present the selection criteria, in Sec.~\ref{Sec:results} we present the analysis of the GC candidates, in Sec.~\ref{sec:azim} we compare the results with the profile of NGC~4709, in Sec.~\ref{sec:LF} we derive the GC luminosity function, in Sec.~\ref{sec:distribution} we present the properties of the GC distribution, and in Sec.~\ref{sec:conclusion} we provide the conclusions.

\section{Observations and data reduction}
\subsection{Description of the data}
\label{sec:data} 

\begin{table}
\caption{Observational properties of the galaxies NGC~4709 and NGC~4696.} 
\centering
\resizebox{\columnwidth}{!}{\renewcommand{\arraystretch}{1.5}\begin{tabular}{c c c c c c c c c c c}

\hline
    Name & RA & DEC & Type & $v_{\rm hel}$ & PA & $e$ & $d$  \\
         & (J2000) & (J2000) & (RC3) & (km s$^{-1}$) & ($^{\circ}$) & & (Mpc)   \\
    \hline
    NGC~4709 & $12^{\rm h}50^{\rm m}03^{\rm s}.88$ & $-41^{\circ}22'55''.10$ & E1 & $4678\pm 4$ & $115$ & $0.167$ & $31.6\pm 2.2$ \\
    NGC~4696 & $12^{\mathrm{h}}48^{\mathrm{m}}49^{\mathrm{s}}.8$ & $-41^{\circ}18'39''$ & cDpec & $2969\pm 12$ & $91$ & $0.28$ & $38.36\pm 2.49$ \\
\hline
\end{tabular}}
\tablefoot{The columns report the name of the object, the coordinates (RA and DEC), the type, and the heliocentric velocity (\citealt{deV}). The ellipticity ($e$) and position angle (PA) were estimated by \citet{Lauberts89}. The distance ($d$) of NGC~4709 was estimated by \citet{mieske} using the Surface Brightness Fluctuation method, whereas that of NGC~4696 was estimated by \citet{Federle} using the globular cluster luminosity function.}
\label{tab:galaxy}
\end{table}

\begin{table}
\caption{Observation summary.}
     
    \label{tab:log}
    \centering
    \resizebox{\columnwidth}{!}{\renewcommand{\arraystretch}{1.5}\begin{tabular}{c c c c c c c}
    \hline
     Date & $T_{{\rm exp},g'}$ & $T_{{\rm exp},r'}$ & $T_{{\rm exp},i'}$ & FWHM$_{g'}$ & FWHM$_{r'}$ & FWHM$_{i'}$  \\
     $\rm{[dd/mm/yy]}$ & (s) & (s) & (s) & (arcsec) & (arcsec) & (arcsec) \\
    \hline
    $21-22/04/2018$ & $12300$ & $11100$ & $14700$ & $0.64$ & $0.6$ & $0.54$ \\
    \hline    
    \end{tabular}}
    \tablefoot{
     The columns report the date of the observation, the exposure times, and the full width at half maximum (FWHM)of the point spread function (PSF) for the different filters.
     }
\end{table}

\begin{figure*}
    \sidecaption
    \includegraphics[width=12cm]{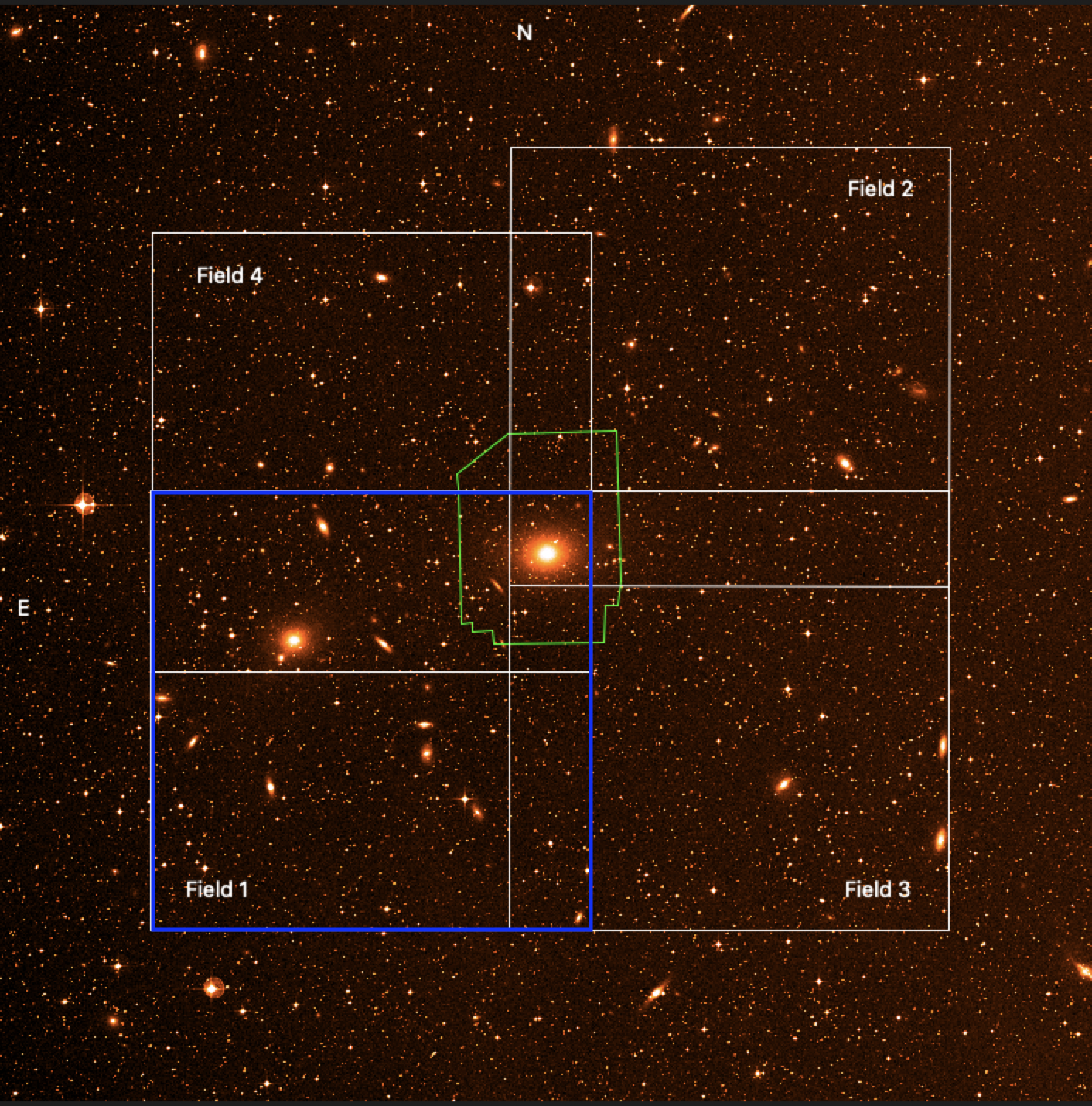}
    \caption{Centaurus cluster. The image shows the four fields obtained during the observation. The field-of-view is of $1\times 1$~deg$^2$, which corresponds to $0.742\times 0.742$~Mpc$^2$, with the north pointing in the upward direction and the east in the leftward direction. The green line delimits the region of $\sim 8.795\times 11.494$~arcmin$^2$ we analyzed in \citet{Federle}, whereas the blue line represents the region of $26.91\times 26.40$~arcmin$^2$ analyzed in this work (Field 1).}
    \label{fig:Centaurus}
\end{figure*}

We observed four fields of the Centaurus cluster (Fig.~\ref{fig:Centaurus}), with MegaCam mounted on the \textit{Magellan} $6.5$ meter telescope during two nights in April 2018. A detailed description of the observations - including field-of-view, data reduction pipeline, exposure times, and image stacking process - is given in \citet{Federle}. Briefly, the observation covers $1\times 1$~deg$^2$ of the Centaurus cluster, which corresponds to $0.742\times0.742$~Mpc. The central giant elliptical galaxy NGC~4696 was located close to the corner of each of the four observed fields.  The data presented here correspond to Field 1 of the Centaurus cluster, which contains the Cen~45 galaxy group, with total exposure times of $3300$, $2400$, and $3600$~s for the $g'$, $r'$, and $i'$ filters, respectively. The field-of-view is of $26.91\times 26.40$~arcmin$^2$ and contains both NGC~4709 and NGC~4696 (see the blue rectangle in Fig.~\ref{fig:Centaurus}). In green in Fig.~\ref{fig:Centaurus}, we overlay the region that we analyzed in \citet{Federle}. We note that the total exposure time is not homogeneous across the indicated field due to the combination of different exposures in order to fill the gaps between CCDs. However, exposure maps were used in our photometry to properly weight each region. The observation log is summarized in Tab.~\ref{tab:log}. 
Most of the data reduction was performed through the dedicated MMT/\textit{Magellan} MegaCam pipeline (\citealt{Megacam15}). Specifically, tasks such as WCS fitting, fringe, and illumination corrections were all done automatically and carefully checked at each reduction step. For this work, we used the individual fits files as processed by the pipeline, but with our own parameters for the stacked mosaic, which was done with Swarp (\citealt{swarp}) and using Astropy (\citealt{astropy1}; \citealt{astropy2}) to perform preliminary statistics on common empty sky regions. The quality and photometric homogeneity of the stacked images were carefully checked in each step. We present here the analysis of Field 1 of the Centaurus cluster, with particular focus on the elliptical galaxy NGC~4709, which is the main component of the Cen~45 group.

\subsection{Subtraction of the galaxy light}
\label{model}
In order to maximize the detection and perform photometry of the GC candidates, we first calculated the models of the surface brightness distribution of NGC~4709 and other four galaxies in Field 1. In this way, the residual images show compact sources otherwise immersed in the galaxies' light. The models of the surface brightness distribution were obtained using {\tt ISOFIT} and {\tt CMODEL} packages (\citealt{ciambur}) in IRAF\footnote{IRAF is distributed by the National Optical Astronomy Observatory (NOAO), directed by the Association of University for Research in Astronomy (AURA) together with the National Science Foundation (NSF). It is available at \url{http://iraf.noao.edu/}}. These are implementations of the {\tt ELLIPSE} and {\tt BMODEL} tasks (\citealt{jed}), which allow the calculation of the model of the surface brightness distribution by fitting elliptical isophotes to the galaxy via a Fourier series expansion. The surface brightness of the isophotes was calculated as:\begin{equation}
    I(\psi)=\langle I_{\rm ell}\rangle + \sum_{n=1}^{\infty}A_n\sin{(n\psi)} + \sum_{n=1}^{\infty}B_n\cos{(n\psi)}
\end{equation}
where $\langle I_{\rm ell}\rangle$ is the surface brightness the isophote would have if it was a perfect ellipse, $\psi$ is the eccentric anomaly, and $A_n$ and $B_n$ are the Fourier coefficients (\citealt{ciambur}).

We used {\tt ISOFIT} because we find it to be more accurate than {\tt ELLIPSE} in the interpolation of the isophotes in this case. In fact, it samples them using the eccentric anomaly, which results in equal length arcs on the ellipse and better describes the deviations from the elliptical form of the isophotes (\citealt{ciambur}). In order to calculate these deviations, the program allows the user to add higher harmonics and use them as input for the {\tt CMODEL} task. We used the first four harmonics, which are given by default, and describe the coordinates of the center of the galaxy, the ellipticity, and the position angle, and, in the case of NGC~4709, we added the $n=6$ harmonics in all three bands. We only used the even harmonics because the odd ones introduce a negligible correction in the case of a galaxy with no significant asymmetries (\citealt{ciambur}), such as NGC~4709. The models were calculated after masking external galaxies and point-like sources present in the field, in order to have a precise measure of the light of the galaxy under consideration. In the case of NGC~4709, to describe the central regions, we left the coordinates of the center, the position angle (PA), and ellipticity as free parameters, whereas in the external regions, we fixed their values in order to reach the faint halo of the galaxy. The parameters used in {\tt ISOFIT} for the analysis of NGC~4709 are listed in Tab.~\ref{tab:isofit properties}. 

\begin{table*}
\caption{Properties of the models of the surface brightness distribution of NGC~4709 in the $i'$ band.}
    \centering
    \begin{tabular}{c c c c c c c c c c c c}
    \hline
    Name & $x_0$ & $y_0$ & PA & $e$ & ${\rm SMA}_0$ & ${\rm SMA}_{\rm max}$ & RA & DEC & $r_{\rm eff}$\\
     & (pixels) & (pixels) & ($^{\circ}$) &  & (arcmin) & (arcmin) & (J2000) & (J2000) & (arcmin)\\ 
    \hline
    NGC~4709 & $4117.31$ & $6923.66$ & $-76.22$ & $0.12$ & $0.719$ & $1.864$ & $12^{\rm h}50^{\rm m}03^{\rm s}.97$ & $-41^{\circ}22'54''.83$ & $0.166$ \\
    \hline
    \end{tabular}
    \tablefoot{The columns show the name of the galaxy, the fixed parameters used to fit the external isophotes ($x_0$, $y_0$, PA, $e$), the values of the semi-major axis between which the isophotes were calculated with fixed parameters (${\rm SMA}_0$ and ${\rm SMA}_{\rm max}$), the coordinates of the center of NGC~4709 corresponding to $x_0$ and $y_0$ (RA and DEC), and the effective radius derived from the model ($r_{\rm eff}$).}
    \label{tab:isofit properties}
\end{table*}

\subsection{Source detection and photometry}
\label{detection}
After the subtraction of the models from the images, we ran the {\tt SExtractor} package (version 2.19.5; \citealt{sex}) to detect sources. After some experimentation, we chose a Gaussian filter, a detection threshold of $3\times \sigma_{\rm sky}$, a detection minimum area of $5$~pixels, a background filter size of $3$, and a background mesh of $24$. Finally, the aperture photometry was performed using an annulus of $2\times FWHM$ and a background annulus of $5$~pixels of thickness. We found $36232$ sources in common in the three filters in Field 1. The photometric calibration was performed using standard stars from the NOAO Source Catalog (\citealt{nsc}). The Galactic extinction varies across the considered field-of-view. To take this into account, we calculated the extinction coefficients for each source using the extinction maps by \citet{SF2011}\footnote{Available at \url{https://irsa.ipac.caltech.edu/applications/DUST/}}. We found $20532$ sources in common between our catalog and the NSC catalog. For the calibration, we considered only those sources for which in the NSC catalog the number of observations was $>3$, and that have a stellarity index $>0.9$ in order to exclude extended sources. Moreover, we visually checked the remaining objects and selected a subsample of $568$ sources spread all across the field for calibration. The magnitudes reported in this work are in the AB system. 

 At the distance $d=42.5\pm 3.2$~Mpc of Centaurus ($m-M=33.14\pm 0.17$~mag; \citealt{mieske}), GCs appear as unresolved sources. In fact, their typical effective radius of $r_{\rm eff}\sim 3$~pc corresponds to just $\sim 0.015$~arcsec here. The analysis of the shape of the sources is then fundamental for a preliminary selection of GCs. {\tt SExtractor} allows the user to estimate the type of object analyzed with the {\tt CLASS\_STAR} parameter, which has a value between zero and one (\citealt{sex}). In particular, objects with ${\tt CLASS\_STAR} = 1$ are stellar (point-like) objects, while those with ${\tt CLASS\_STAR} = 0$ are extended sources. In order to separate GCs from galaxies, we divided the detected sources into two subsamples with ${\tt CLASS\_STAR} > 0.5$ and ${\tt CLASS\_STAR} \leq 0.5$ in the $i'_0$ filter, which corresponds to the image with the best seeing. In this way, we found $18651$ compact (or unresolved) sources with ${\tt CLASS\_STAR} > 0.5$, which are shown in the color-magnitude diagram in the top panel of Fig.~\ref{fig:cmd}.

\section{Selection of globular cluster candidates}
\label{sec:selection}

\begin{figure}
    \centering
    \includegraphics[scale=0.3]{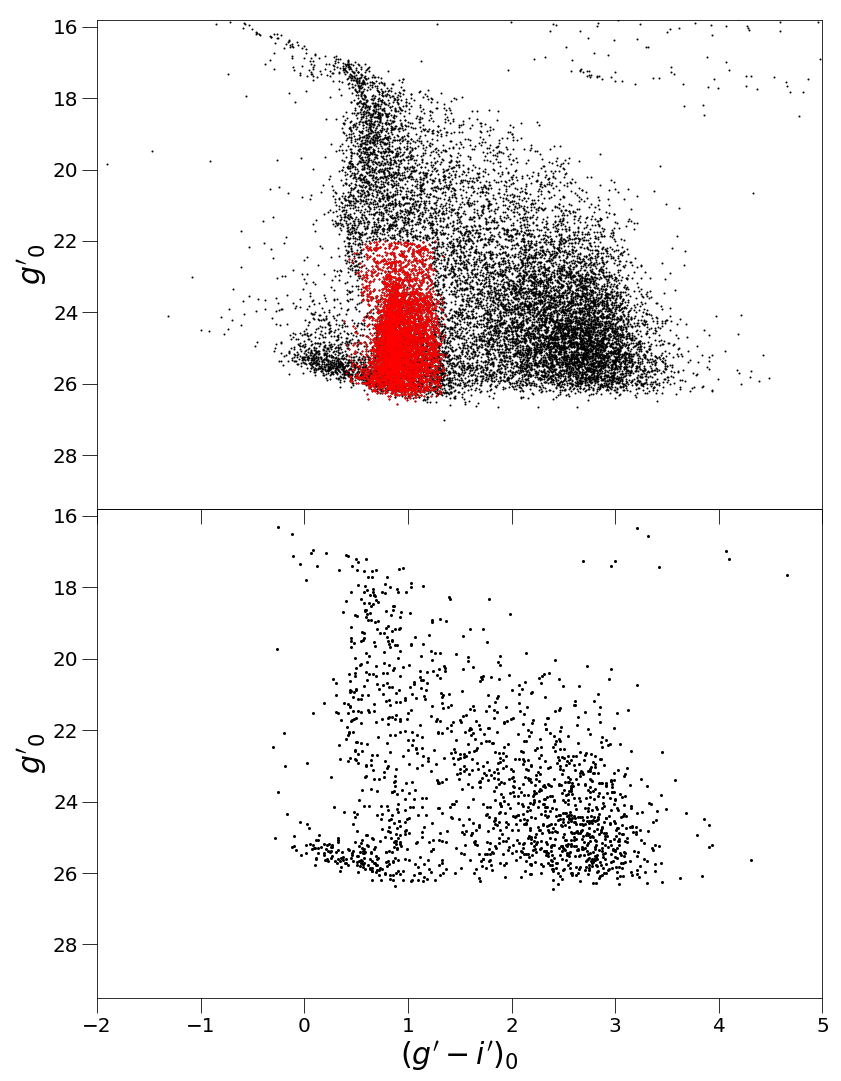}
    \caption{Comparison between the color magnitude diagram of the target field and that of the background region. {\em{Top panel}}: Color-magnitude diagram of the $18651$ sources (black dots) selected according to their shape. The plot shows the $g'_0$ magnitude as a function of the $(g'-i')_0$ color. The red dots represent the sources after the selection on colors and magnitudes. {\em{Bottom panel}}: Color-magnitude diagram of the background region. The plot shows the $g'_0$ magnitude as a function of the $(g'-i')_0$ color for the $1639$ compact sources ($8.79\%$ of the total sources in Field 1) located in the rectangle region of $11.10\times 6.95$~arcmin$^2$ ($11.87\%$ of Field 1) used to calculate the background level (see green rectangle in Fig.~\ref{fig:positions}).}
    \label{fig:cmd}
\end{figure}

\begin{figure}
    \centering
    \includegraphics[scale=0.27]{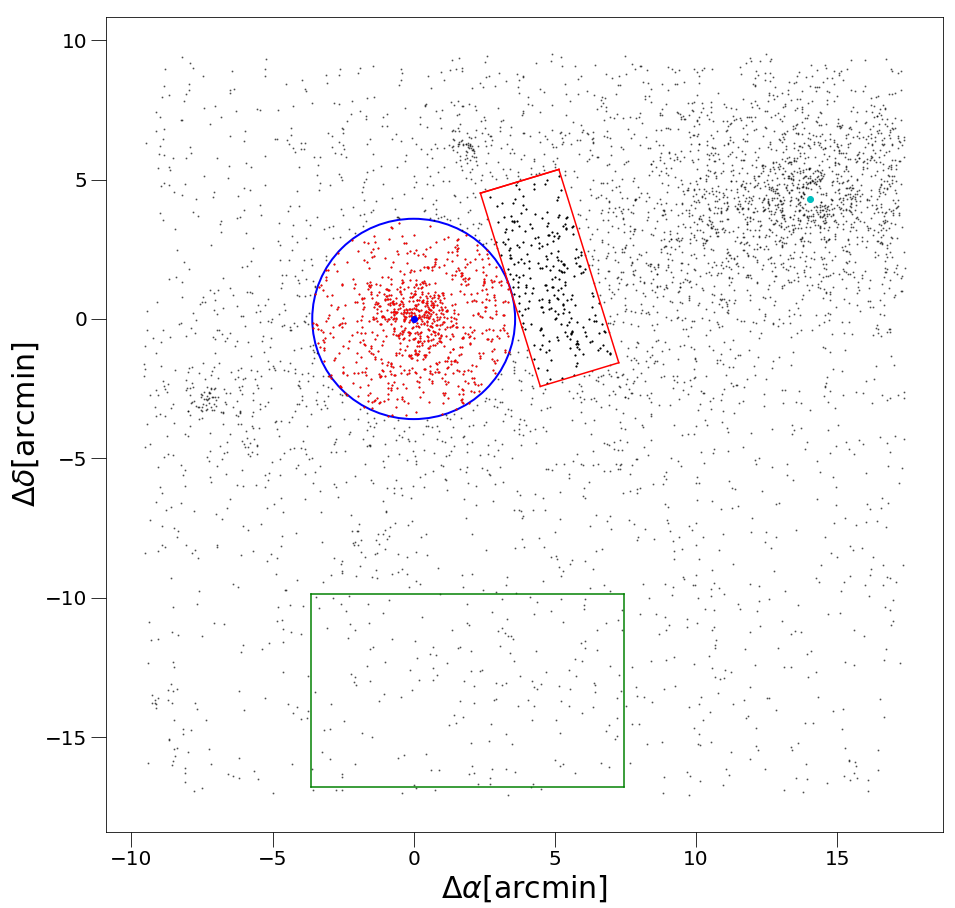}
    \caption{Positions of the GC candidates with respect to the center of the galaxy. The plot shows the declination and right ascension of the sources, where the center of NGC~4709 (blue point) is located at $(\Delta\delta, \Delta\alpha)=(0,0)$. The blue circle defines the limit used to define the GCS of NGC~4709. The green rectangle represents the $\sim 77.22$~arcmin$^2$ region where the contamination level from background objects was calculated, the cyan point is the position of the center of NGC~4696, and the red rectangle is the $\sim 20.89$~arcmin$^2$ region between NGC~4709 and NGC~4696 for which we calculated the luminosity function in Sec.~\ref{sec:distribution}. The field-of-view is of $26.91\times 26.40$~arcmin$^2$.}
    \label{fig:positions}
\end{figure}

Figure~\ref{fig:cmd} shows a very large spread in color of the selected sources. In order to eliminate potential contaminants from the $18651$ compact sources, such as foreground stars and background objects, we applied a second selection criterion based on $(g'-r')_0$ and $(r'-i')_0$ colors, and on the $g'_0$ magnitudes. In particular, we used $0.35<(g'-r')_0 < 0.85$, $0.0 < (r'-i')_0 < 0.55$, and $g'_0 > 22.0$ as described in \citet{Faifer17}. These limits were obtained in \citet{Faifer11} for the GCSs of five early-type galaxies located at a distance modulus between $29.93\pm 0.09$ and $31.90\pm 0.20$. Given these distances and that of NGC~4709 (see Tab.~\ref{tab:galaxy}), we could apply the same color selection since redshift effects are negligible. The color limits were obtained by \citet{Faifer11} by considering the completeness of the GC sample and the magnitudes of Galactic GCs. In particular, the bright limit corresponds to $M_I > -12$~mag, which represents the typical separation between GCs and ultra-compact dwarfs (UCDs). Moreover, as shown in Fig.~8 in \citet{Faifer11}, the limits cover the metallicity range expected for GCs. Our selection considers a somewhat larger range in color to take into account for the increasing errors toward fainter magnitudes. Using these limits, we obtained a sample of $5191$ GC candidates, whose properties are summarized in Tab.~\ref{tab:appendix}. From Fig.~\ref{fig:cmd}, we observed that there are a lot of unselected sources (black dots) in the region defined by our selection. This is caused by the fact that the selection was made only on the $(g'-r')_0$ and the $(r'-i')_0$ colors and not on the $(g'-i')_0$ color. Fig.~\ref{fig:positions} shows the positions of the sources with respect to the center of NGC~4709.

\section{Results}
\label{Sec:results}

\subsection{Color-magnitude and color-color diagrams}
\label{sec:cmd}

\begin{figure}
    \centering
    \includegraphics[scale=0.3]{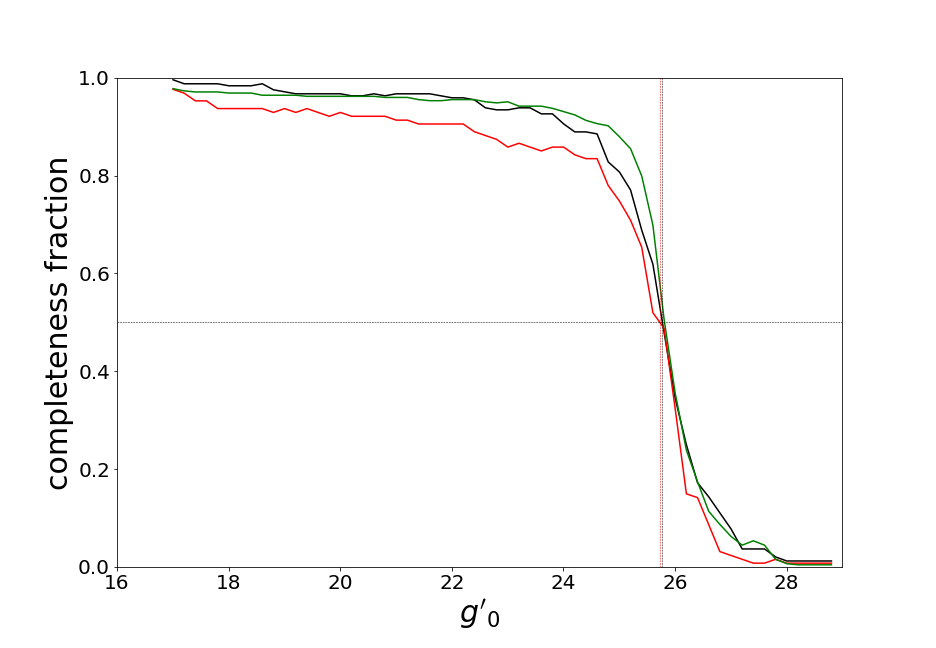}
    \caption{Completeness test. The plot shows the completeness fraction as a function of the $g'_0$ magnitude for the artificial stars added in the three filters inside a galactocentric radius of $5\times r_{\rm eff}$ (black line), for the artificial stars inside the rectangular region used to estimate the background contamination (green line), and inside the region between NGC~4709 and NGC~4696 (red line) represented by the red rectangle in Fig.~\ref{fig:positions}. The dashed lines show the $50\%$ completeness limit.}
    \label{fig:completeness}
\end{figure}

At this point, the sample of GC candidates defined using a selection on the stellarity index (${\tt CLASS\_STAR} > 0.5$ in the $i'_0$ filter), the colors ($0.35 < (g'-r')_0 < 0.85$ and $0.0 < (r'-i')_0 < 0.55$), and magnitude ($g'_0 > 22$), includes $5191$ sources. In order to test the detection limits of our photometry, we performed the completeness test of the GC sample. In particular, we used the {\tt ADDSTAR} task (\citealt{Stetson87}) in {\tt IRAF} to add $5000$ artificial star-like objects per each magnitude considered based on the PSF estimated with {\tt DAOPHOT} (\citealt{Stetson87}). The task allows the user to add point-like sources at chosen magnitudes and positions, which in our case is important to recover their colors. The artificial stars were then added in the three filters at chosen positions in steps of $0.2$ magnitudes, so that the total number of sources added with magnitudes of $17.0 \leq g'_0 \leq 28.8$~mag is $300000$. Given the large field-of-view, the addition of $5000$ objects in each step allowed us to get a good number statistics, but also to avoid crowding effects. Using {\tt SExtractor} with the same parameters as before, we determined how many of the added stars could be recovered in the three filters. The magnitudes were corrected using the same zeropoint as before. Moreover, we made sure that the artificial stars were added in the same positions in all three filters, and that they had a color of $(g'-i')_0 = 0.878$, which represents the mean for our sample. Finally, we limited our completeness test to a circular region around NGC~4709 with a galactocentric radius of $r = 3.593$~arcmin, which corresponds to $5\times r_{\rm eff}$. In this region, we have $244$ artificial stars and $881$ GC candidates. In Fig.~\ref{fig:completeness}, the $50\%$ completeness limit is reached at a magnitude $g'_0 = 25.77$~mag, so by choosing to limit our study at $g'_0 \leq 25.46$~mag, where we have a $67\%$ completeness (as we did in \citealt{Federle}), we are well above the $50\%$ completeness.

Despite the morphological and color selection criteria, some of the point-like sources are likely to be foreground or background objects, such as Milky Way stars or background galaxies, so an additional correction must be applied. For an upper limit estimate of the potential background contamination, we selected a rectangular region with an area of $\sim 77.22$~arcmin$^2$ located in the lower part of the image at a sufficiently large distance from NGC~4709, with the upper limit of the rectangle at a galactocentric distance of $\sim 13.6\times r_{\rm eff}$ (see the green region in Fig.~\ref{fig:positions}). We ran the photometry and completeness test on the rectangular region using the same parameters as before. The total number of sources in the background region is $3432$, of which $1639$ were classified as compact by {\tt SExtractor}. The number of sources after the shape, colors, and magnitude selection is $169$. The color-magnitude diagram of the sources in the background region is shown in the bottom panel of Fig.~\ref{fig:cmd}. Of the sources added with the {\tt ADDSTAR} task, $448$ fall inside the rectangular region. The results of the completeness test for the background region are shown by the green line in Fig.~\ref{fig:completeness}. In order to correct the number of GC candidates for the background level, we corrected the background counts for the completeness fraction in the rectangular region and rescaled them to take into account the difference between the area under consideration and that in which the background was estimated. After the correction, we have a total number of $88$ background sources. We divided both the sources in the background region and the GC candidates in bins of $0.3$~mag according to their $(g'-i')_0$ color, calculated the number of sources and background in each bin, and used the function {\tt random.sample()} in Python to randomly discard the same number of background sources in each bin from our GC sample. The final number of GC candidates of NGC~4709 located inside a galactocentric radius of $3.593$~arcmin ($5\times r_{\rm eff}$; blue circle in Fig.~\ref{fig:positions}) reduces to $556$.

Finally, to estimate which fraction of sources corresponds to foreground stars in the Milky Way, we compared our results with the star counts given by the Besan{\c{c}}on model of the Galaxy\footnote{Available at \url{http://model.obs-besancon.fr}} (\citealt{Robin03}). We ran the simulation for the MegaCam filters, considering a region with the same area as our background region and with $18.0\leq g'_0\leq 28.0$~mag. With these values, we found $17$ sources, of which only two are within the range of the color-magnitude cut applied to select our GC candidates. From this result, we concluded that the contribution of foreground stars to the background level is negligible. 

\subsection{Color distribution}
\label{sec:color distribution}

\begin{figure*}
    \centering
    \includegraphics[width=\textwidth]{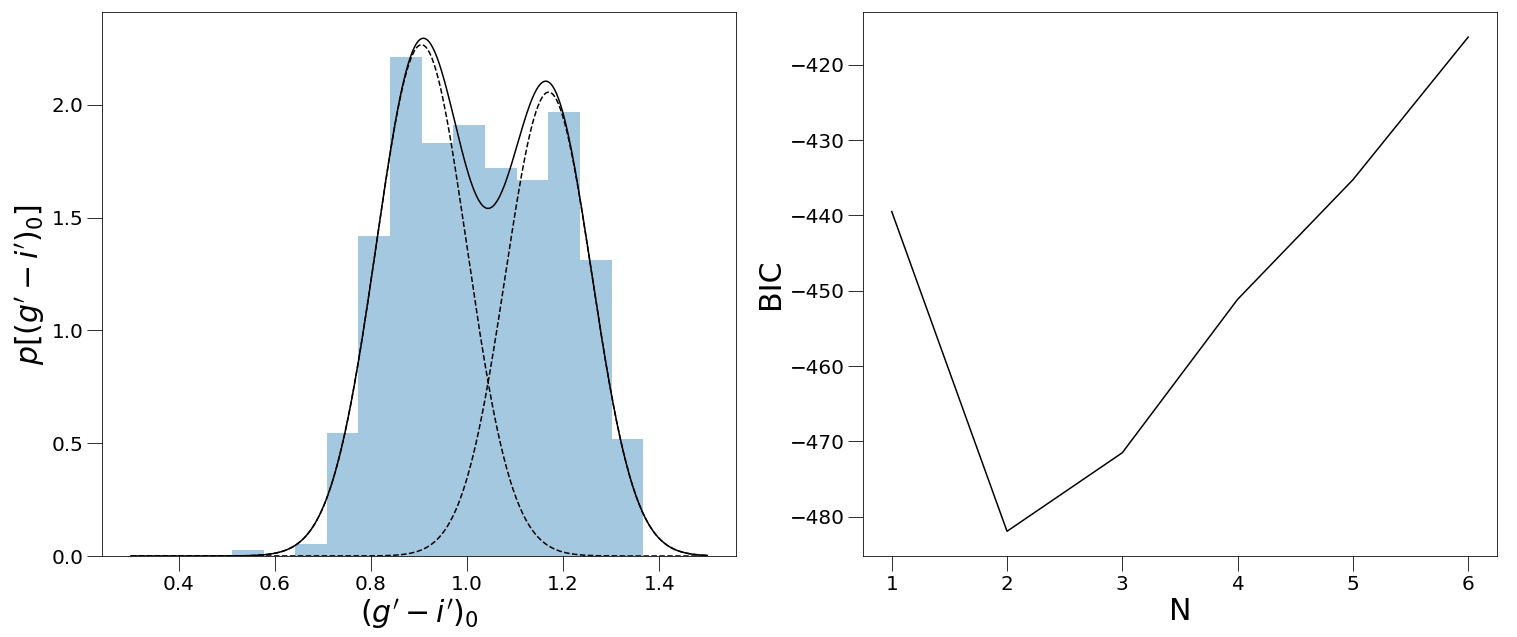}
    \caption{Bayesian Information Criterion test on the GCS of NGC~4709. {\em Left panel}: Probability density function versus $(g'-i')_0$ color. The black line represents the best-fit model, the dashed lines represent the two Gaussians described by the model. The color distribution is divided in bins according to the Freedman-Diaconis rule (\citealt{Freedman81}). {\em Right panel}: Values of the BIC as a function of the number of components in the model.}
    \label{fig:BIC}
\end{figure*}

The color distribution of the $556$ GC candidates, selected according to the selection criteria in Sec.~\ref{sec:selection} and corrected for the background level, is more complex than a single Gaussian. Bimodality has been found in most GCSs of elliptical galaxies and has been confirmed in NGC~4696, the central giant elliptical of the Centaurus cluster (\citealt{Federle}). To investigate this issue in NGC~4709 we analyzed the color distribution of the GC candidates using the Gaussian Mixture Modeling (GMM) of {\tt sklearn} in Python (\citealt{Pedregosa11}), which allows to fit multiple Gaussians to check for multimodality. From this, we found that a bimodal Gaussian, with the blue peak at $(g'-i')_0 = 0.905\pm 0.009$~mag and the red peak at $(g'-i')_0 = 1.170\pm 0.008$~mag, is preferred over a unimodal or a trimodal one (see Fig.~\ref{fig:BIC}). The red and blue populations were divided at the color for which the GMM fit gave an equal probability for the object of belonging to the red or blue subsamples (as described in \citealt{Escudero18}). In our case, the color cut of the total sample corresponds to $(g'-i')_0 = 1.044$~mag, which is redder than what we found for the GCS of NGC~4696 (\citealt{Federle}). To test whether two Gaussians are preferable instead of three or more, we used the Bayesian Information Criterion (BIC; \citealt{Schwarz}) implemented in astroML (\citealt{astroML}), which, given the data and a certain number of components for the model, estimates the quality of the model. We performed the BIC test using a number of components from one to six. As can be seen from Fig.~\ref{fig:BIC}, the best-fit, for which the value of the BIC is minimum, is obtained for a model with two Gaussians.

\section{The shape of the globular cluster system and comparison with NGC~4709 profile}
\label{sec:azim}

We studied the angular distribution of the GC candidates around NGC~4709, since this is an additional test to probe its interaction history. In particular, the comparison with the position of NGC~4696 is important since \citet{Walker13} outlined that the X-ray emission shows signs of interactions between these galaxies, and in \citet{Federle} we showed that the GCS of NGC~4696 has clear peaks in the directions of NGC~4709 and NGC~4696B. 

The azimuthal distribution is described by an asymmetric sinusoidal, with peaks at position angles of PA$_1\sim 92.19^{\circ~+2.36}_{~-4.91}$ and PA$_2\sim 293.47^{\circ~+1.08}_{~-2.57}$. PA$_2$ coincides within the errors with the direction of NGC~4696, confirming that the interaction between the two galaxies shaped not only the GCS of NGC~4696 as we found in \citet{Federle}, but also the GCS of NGC~4709. 

The sinusoidal shape of the GCS is frequently found among elliptical galaxies. In order to calculate the parameters of the distribution of the GCS, we performed the 2D Kernel Density Estimation of {\tt sklearn.neighbors} (\citealt{Pedregosa11}) to create a grid of the sample locations using the physical positions of the GC candidates in the image, and we used the results to fit a Bivariate Gaussian Distribution using a routine in astroML (\citealt{astroML}). The outputs of this routine are the position of the center of the ellipse, the sigmas of the two Gaussians, which can be used as major and minor axes to calculate the ellipticity and the position angle. The results within a $68\%$ confidence interval are shown in Tab.~\ref{tab:shape}. 

NGC~4709 is an elliptical galaxy of type E1 (\citealt{deV}). We ran {\tt ISOFIT} using the Fourier coefficients $a_3$, $a_4$, $a_6$, $b_3$, $b_4$, and $b_6$, to quantify the deviations from the elliptical shape of the isophotes (\citealt{ciambur}). Our model is limited to a galactocentric radius of $1.864$~arcmin, where the sky level is reached. From our results, we found an effective radius of $~0.719$~arcmin and an ellipticity of $e=0.121\pm 0.02$, which is lower than the values of $e=0.255^{+0.008}_{-0.010}$, $e=0.294^{+0.029}_{-0.021}$, and $e=0.257^{+0.010}_{-0.013}$ found for the total, red and blue GC populations. We found a position angle of PA$=103.78\pm 0.57$ for NGC~4709, which is different than what was found for the GCS. In particular, the GCs distribution shows peaks at the position angles of ${\mathrm{PA}}=92.19^{+2.36}_{-4.92}$~deg and ${\mathrm{PA}}=293.47^{+1.08}_{-2.56}$~deg. This means that the GC distribution is not aligned with the radial profile of the host galaxy, which points toward a disturbed GCS, with a shape that was influenced by past interactions with NGC~4696. Our results, together with what we found for NGC~4696 in \citet{Federle}, agree with the interaction scenario highlighted by X-ray studies that found a metallicity excess around NGC~4709 (\citealt{Walker13}) and by asymmetric temperature variations in the X-ray gas with the hottest regions coinciding with NGC~4709 (\citealt{Churazov99}).

\begin{table*}
\caption{Shape of the GCS}
    \centering
    \begin{tabular}{c c c c c c c c}
    \hline
    & $N_{\rm GC}$ & RA & DEC & $a$ & $b$ & $e$  \\
    & & (J2000) & (J2000) & (arcmin) & (arcmin) & \\
    \hline
    Total & $556$ & $12^{\rm h}50^{\rm m}04^{\rm s}.35^{+0.11}_{-0.10}$ & $-41^{\circ}22'58''.17^{+3.25}_{-2.16}$ & $3.37^{+0.02}_{-0.03}$ & $2.51^{+0.04}_{-0.03}$ & $0.256^{+0.012}_{-0.013}$ \\
    Red & $254$ & $12^{\rm h}50^{\rm m}04^{\rm s}.269^{+0.004}_{-0.272}$ & $-41^{\circ}22'51''.72^{+6.13}_{-6.79}$ & $3.27^{+0.06}_{-0.07}$ & $2.31^{+0.06}_{-0.10}$ & $0.294^{+0.029}_{-0.023}$ \\
    Blue & $302$ & $12^{\rm h}50^{\rm m}04^{\rm s}.37^{+0.14}_{-0.11}$ & $-41^{\circ}23'01''.18^{+6.27}_{-9.21}$ & $3.36^{0.03}_{-0.04}$ & $2.47^{+0.08}_{-0.11}$ & $0.265^{+0.035}_{-0.025}$ \\
    \hline
    \end{tabular}
    \tablefoot{The columns show the coordinates of the center of the system (RA and DEC), the major and minor axis ($a$ and $b$), and the ellipticity ($e$) for the total, red and blue GC populations.}
    \label{tab:shape}
\end{table*}

\section{Globular cluster luminosity function and specific frequency}
\label{sec:LF}

The globular cluster luminosity function (GCLF) is an important tool to determine the richness of a GCS. Furthermore, its turnover magnitude is used as a distance estimator for its host galaxy (e.g., \citealt{Rejkuba12}). It is defined as:\begin{equation}
    \frac{\mathrm{d}N}{\mathrm{d}M}\propto \frac{1}{\sigma\sqrt{2\pi}}\exp\left[-\frac{(m-M)^2}{2\sigma^2}\right]
    \label{eq:LF}
\end{equation}
where $\mathrm{d}N$ is the number of GCs in the magnitude bin $\mathrm{d}m$, $m$ and $M$ are the apparent and absolute turnover magnitudes, and $\sigma$ is the width of the Gaussian distribution.

In this section, we estimate the distance to NGC~4709 from the GCLF and, based on that distance, we determine the specific frequency, $S_N$, of the GCS. $S_N$ is related to the total number of GCs and to the luminosity of the host galaxy in the $V$ band in units of $M_V=-15$, and it is given by (\citealt{Harris81}):\begin{equation}
    S_N = 10^{0.4\cdot(M_V+15)} N_{\rm GC}
    \label{eq:SF}
\end{equation}
where $N_{\rm GC}$ is the total number of GCs and $M_V$ is the absolute magnitude of the host galaxy. It has been shown that the specific frequency depends on the galaxy type and luminosity, with the highest specific frequencies occurring in low-mass dwarf galaxies and in giant ellipticals in galaxy clusters, where it reaches values of $\sim 10$ for $M_V<-20$~mag (\citealt{Peng08}; \citealt{Georgiev10}).
We assumed the absolute turnover magnitude in the three bands to be $M_{g'}=-7.20$~mag, $M_{r'}=-7.56$~mag, and $M_{i'}=-7.82$~mag (\citealt{Jordan07}). 

As explained in Sec.~\ref{sec:cmd}, we did not apply any statistical correction for foreground stars since the Besan{\c{c}}on Model of the Galaxy (\citealt{Robin03}) shows that the contribution of foreground stars in this region is negligible. We calculated the contamination from foreground and background sources from a $\sim 77.22$~arcmin$^2$ rectangular region located in the lower part of the image (see Fig.~\ref{fig:positions}), with the upper side at a galactocentric distance of $\sim 13.6\times r_{\rm eff}$. After the shape, color, and magnitude selection, the number of sources in this region is $169$. We divided our sample in magnitude bins of $0.2$~mag and corrected the number of GCs in the three bands, such as:\begin{equation}
    N_{\rm GC}=\frac{N}{f}-A_{\rm fac}\frac{N_{\rm bkg}}{f_{\rm bkg}}~,
\end{equation}
where $N$ is the number of GC candidates after our selection (see Sec.~\ref{sec:selection}), $f$ is the completeness fraction, $N_{\rm bkg}$ is the number of background sources, $f_{\rm bkg}$ is the completeness fraction in the background region, and $A_{\rm fac}$ is a factor that takes into account the difference between the area of the GCS and the background area. 

\begin{figure}
    \centering
    \includegraphics[scale=0.27]{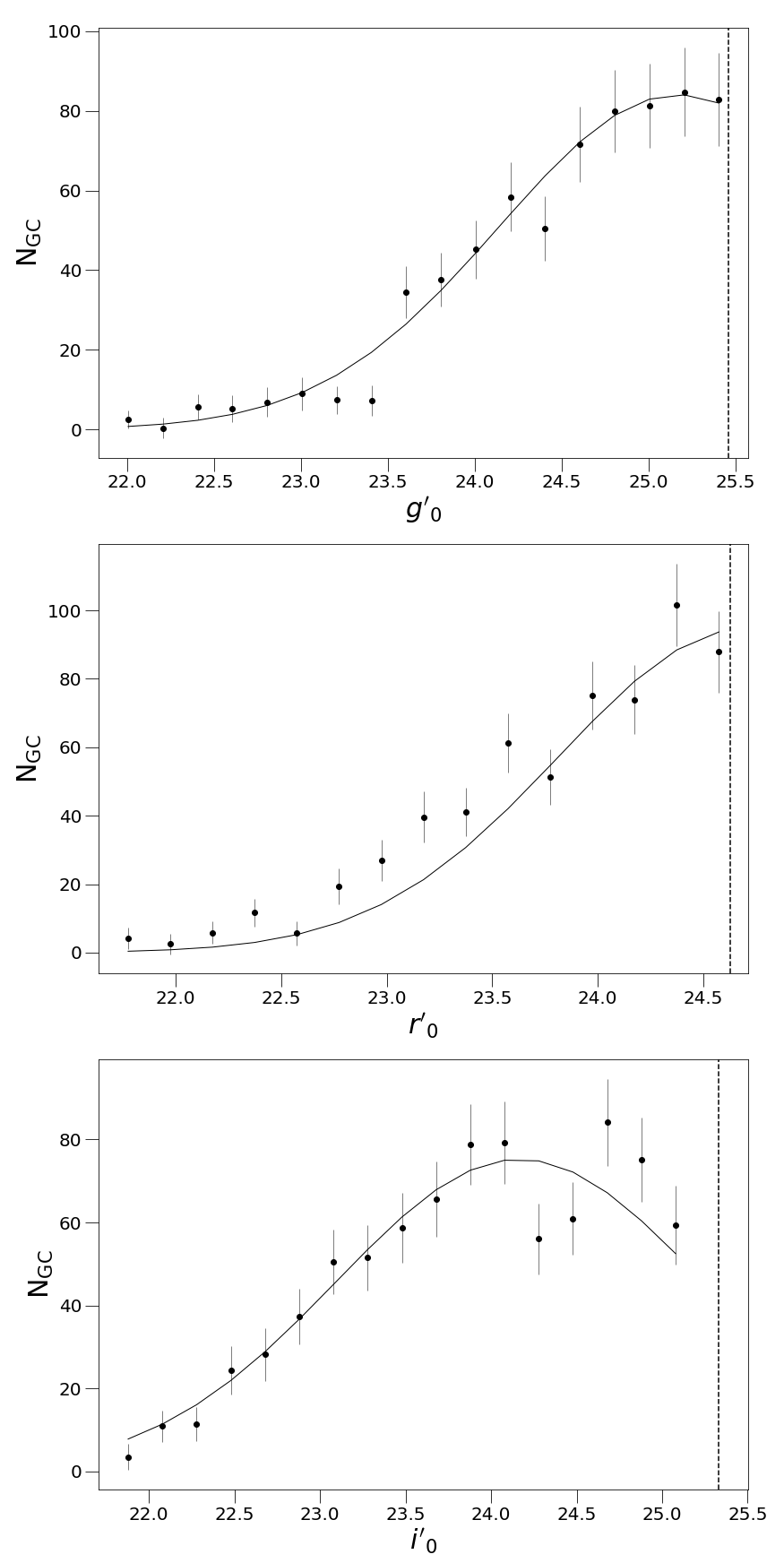}
    \caption{Globular cluster luminosity function. {\em Top, middle, and bottom panels}: GCLF in the $g'_0$, $r'_0$, and $i'_0$ bands, where the black points represent the number of clusters corrected for the completeness fraction and for the background level divided in bins of $0.2$~mag, and the solid line represents the best-fit Gaussian. The dashed lines show the magnitudes at which we have $67\%$ completeness.}
    \label{fig:LF}
\end{figure}

The results of the Gaussian fit in the $g'$, $r'$, and $i'$ bands are shown in Fig.~\ref{fig:LF}, whereas the magnitudes up to which the fit was performed, the best-fit parameters, and the calculated distances are reported in Tab.~\ref{tab:LF}. The derived distances in the three bands agree with each other within their error bars. The error weighted mean distance between the three bands is $29.9\pm 2.1$~Mpc, which is consistent with both the distance of $31.6\pm 2.2$~Mpc obtained with the surface brightness fluctuation method and that of $29.1\pm 4.4$~Mpc obtained through the study of the GCLF by \citet{mieske} and by \citet{Mieske03}. In \citet{Federle} we calculated the distance of the central giant elliptical NGC~4696, which resulted of $38.36\pm 2.49$~Mpc. This value is again in good agreement with the distance obtained by \citet{mieske} using both the surface brightness fluctuation method and the GCLF. With our data, we therefore derived a relative distance of $8.46\pm 3.26$~Mpc between NGC~4709 and NGC~4696, which is similar to what was found by \citet{mieske} and places NGC~4709 in front of the main Centaurus cluster. However, we note that the error bars of both the distances of NGC~4709 and NGC~4696 are quite large, so the distance between the two galaxies could be much lower than what we estimate in this work. Despite the differences in the radial velocities of the two components of Centaurus (\citealt{Lucey86}), X-ray analysis of the intracluster medium showed that Cen~45 and Cen~30 are located at the same redshift of $z\sim 0.0104$ (\citealt{Ota16}), which would be consistent with the distance between NGC~4709 and NGC~4696 being in the lower range given by the error bar. Nevertheless, most of the main galaxies have distances $>42$~Mpc (see Tab.~5 in \citealt{mieske}), which makes NGC~4709 an outlier in the Centaurus cluster. 

Besides the distance, the GCLF can be used to estimate the total number of GCs in the system, which is important to determine the richness of the GCS. The total number of clusters is given by \begin{equation}
    N_{\rm GC}=\frac{\sqrt{2\pi}A \cdot \sigma}{\mathrm{bin ~size}}
\end{equation}
where $A$ and $\sigma$ are the amplitude and standard deviation of the GCLF, and the ${\mathrm{bin~size}}=0.2$~mag. The total number of sources was obtained by integrating the GCLF, resulting in $N_{{\rm GC},~g'}=1105$, $N_{{\rm GC},~r'}=1331$, and $N_{{\rm GC},~i'}=1142$, respectively. We adopted the mean of those three numbers $N_{\rm GC}=1193$. As the apparent magnitude for NGC~4709, we assumed the value of $m_V=11.1$, found by \citet{Misgeld09}. Assuming a distance modulus of $(m-M)=32.38\pm 0.15$ ($29.9$~Mpc as determined previously), we found $M_{V,~{\rm NGC~4709}}=-21.28$~mag. With these values, we obtained a specific frequency of $3.7\pm 0.5$. This agrees within the error bars with the value of $5.0\pm 1.3$ found by \citet{mieske}, and with typical values obtained for other elliptical galaxies.

\begin{table*}
    \caption{Best-fit parameters for the GCLF of NGC~4709.}
    \centering
    \begin{tabular}{c c c c c c c c}
    \hline
    Band & mag & $A$ & $\mu$ & $\sigma$ & $M_{\rm TO}$ & $d$ & $(m-M)$\\
         & (mag) & & (mag) & (mag) & (mag) & (Mpc) & (mag) \\
    \hline
    $g'_0$ & $25.457$ &  $84.08\pm 0.09$ & $25.22\pm 0.02$ & $1.05\pm 0.02$ & $-7.20\pm 0.20$ & $30.49\pm 3.35$ & $32.42\pm 0.24$ \\
    $r'_0$ & $24.627$ & $94.77^{+1.67}_{-1.66}$ & $24.79\pm 0.05$ & $1.12\pm 0.02$ & $-7.56\pm 0.20$ & $29.46\pm 5.04$ & $32.35\pm 0.37$ \\
    $i'_0$ & $25.332$ & $74.92^{+0.82}_{-0.78}$ & $24.53\pm 0.02$ & $1.22\pm 0.02$ & $-7.82\pm 0.20$ & $29.49\pm 3.18$ & $32.35\pm 0.23$\\
    \hline
    \end{tabular}
    \tablefoot{The columns show the band, the magnitude up to which the LF was fitted (mag), the amplitude ($A$), the mean ($\mu$), and the standard deviation ($\sigma$). The last three columns show the absolute turnover magnitude for the filter considered, and the calculated distance and distance modulus for NGC~4709.}
    \label{tab:LF}
\end{table*}

\section{Properties of the globular cluster distribution}
\label{sec:distribution}

\begin{figure}
    \centering
    \includegraphics[scale=0.27]{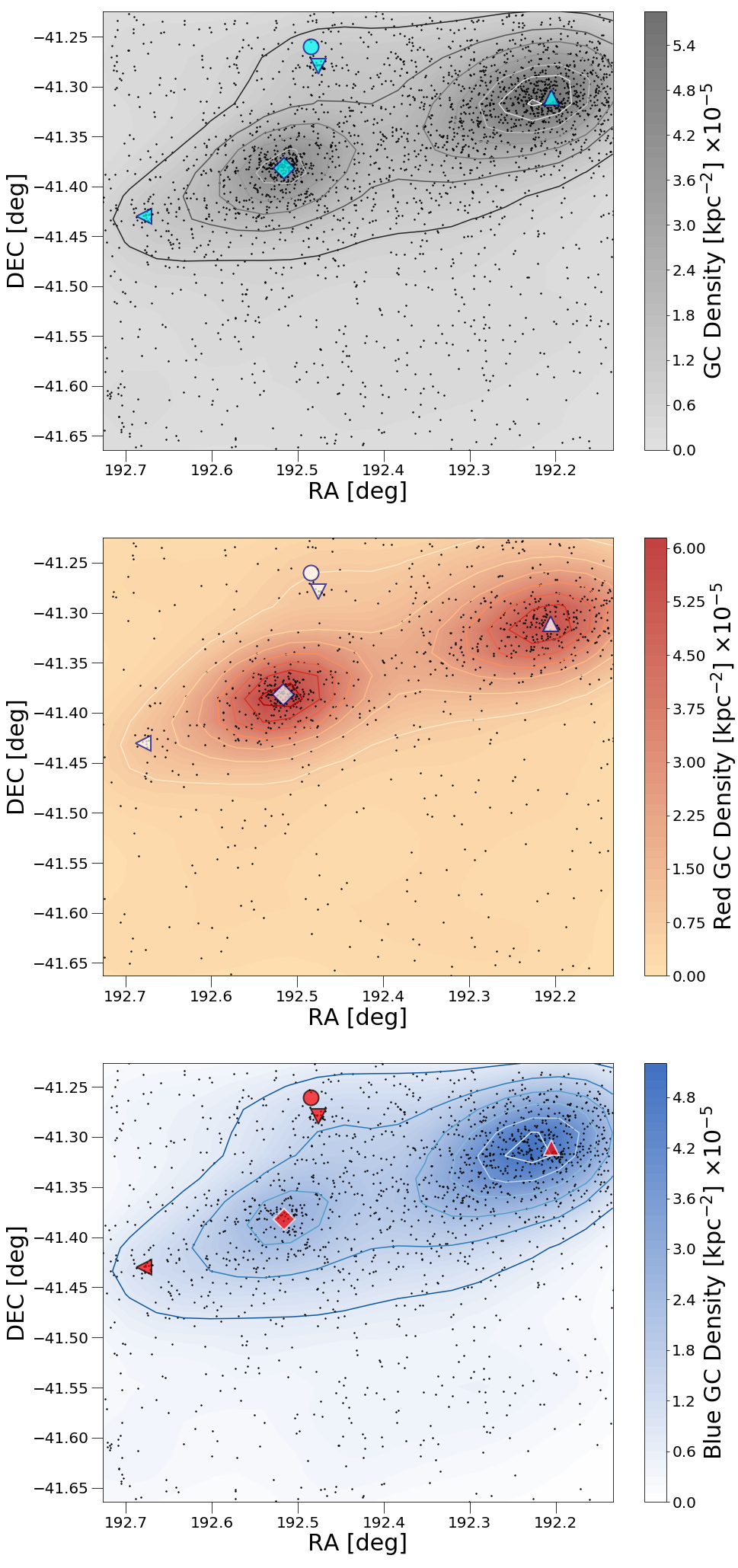}
    \caption{Globular cluster candidates density maps of the full Field 1. The plots show the density maps for the total ({\em top panel}), red ({\em middle panel}), and blue ({\em bottom panel}) GC populations, and the positions of NGC~4709 (diamond), NGC~4696 (triangle), NGC~4706 (inverted triangle), WISEA~J124956.33-411536.8 (circle), and ESO~323-G~009 (leftward triangle). The field-of-view is of $26.91\times 26.40$~arcmin$^2$.}
    \label{fig:density maps}
\end{figure}

The spatial distribution of the GC candidates in the $\sim 26.91\times 26.40$~arcmin$^2$ region of Field 1 is shown by the density maps in Fig.~\ref{fig:density maps}. 
The total number of candidates was corrected for the background value, so the final number of GC candidates is $N_{\rm GC}=2666$. As shown in Fig.~\ref{fig:total color}, the total color distribution in Field 1 is bimodal, with blue and red peaks at a color of $(g'-i')_0=0.856\pm 0.009$~mag and $(g'-i')_0=1.111\pm 0.010$~mag, respectively. The separation between the two populations is at a color of $(g'-i')_0=1.000$~mag. Using this value, we found $N_{\rm red}=965$ and $N_{\rm blue}=1701$ for the red and blue populations, respectively. As can be seen from Fig.~\ref{fig:density maps}, the blue GC population is more evenly distributed throughout Field 1 with respect to the red one, which outside of the main galaxies shows regions with fewer GCs. This is in good agreement with the accreted origin scenario for the blue GC candidates (\citealt{Brodie06}) and with results found for other GCSs (e.g. \citealt{Urbano24}). Besides the regions of NGC~4709 and NGC~4696, we found overdensities of blue GC candidates that coincide with the positions of NGC~4706, WISEA J124956.33-411536.8, and ESO~323-G~009, which were confirmed to be part of the Centaurus cluster (e.g. \citealt{Chiboucas07}). We also note that the blue GCs seem to dominate in the region between NGC~4709 and NGC~4696. To further investigate these findings, we studied the color distribution of the GC candidates in these four galaxies and in the $\sim 20.89$~arcmin$^2$ rectangular region between NGC~4709 and NGC~4696 (see Fig.~\ref{fig:positions}). We limited our analysis to $22.0<g'_0<25.46$~mag and to galactocentric distances of $5\times r_{\rm eff}$, as was done for NGC~4709. In the case of NGC~4696, we used the value for the effective radius reported in the literature ($r_{\rm eff}=1.7$~arcmin) since the value we obtained from our analysis is lower due to the fact that the galaxy is located at the edge of the image so our model does not reach its outer halo. The number of sources, peaks of the color distributions, and color separations between the blue and red populations are reported in Tab.~\ref{tab:color dis galaxies}. The GC candidates of NGC~4696 and ESO~323-G~009 show a bimodal color distribution, whereas those of NGC~4706 have an unimodal distribution. The GCS of WISEA~J124956.33-411536.8 has a very low number of GCs, so it is not possible to calculate a reliable color distribution. The parameters of the four galaxies and of their GCSs are reported in Tab.~\ref{tab:galaxies parameters} and Tab.~\ref{tab:shape GCSs}.

In \citealt{Federle} we determined that the interactions between NGC~4696 and NGC~4709 shaped the GCS of NGC~4696. As seen in Sec.~\ref{sec:LF}, the two galaxies belong to different components of the Centaurus cluster (Cen~30 and Cen~45, respectively) and are located at different distances, with NGC~4709 being $\sim 8.5$~Mpc closer to us than NGC~4696. In order to see if the differences in distances could be reflected by the GC candidates' properties, we analyzed the sources located in a $\sim 20.89$~arcmin$^2$ rectangular region between NGC~4709 and NGC~4696 (red rectangle in Fig.~\ref{fig:positions}). In particular, the long sides of the region are located at a distance of $5\times r_{\rm eff}$ from the two galaxies, which corresponds to $3.593$~arcmin and $8.5$~arcmin for NGC~4709 and NGC~4696, respectively. Using the same selection criteria as in Sec.~\ref{sec:selection}, we found $202$ GC candidates in this region. A comparison between the color-magnitude diagram of these sources and the GC candidates around NGC~4709 is shown in Fig.~\ref{fig:cmd bet}. 

The results of the completeness test in this region are shown by the red curve in Fig.~\ref{fig:completeness}. After correcting the counts for the background level, the number of sources in this region decreased to $109$. We analyzed the color distribution using the GMM of {\tt sklearn} in Python (\citealt{Pedregosa11}), and we used the Akaike Information Criterion (AIC; \citealt{Akaike}) implemented in astroML (\citealt{astroML}) to test whether the distribution is uni- or multimodal. In this case, we preferred the AIC to the BIC due to the small sample size. We performed the AIC for a number of components from one to six. As can be seen from Fig.~\ref{fig:AIC} the best-fit, for which the value of the AIC is minimum, is obtained for a model with two Gaussians, with peaks at $(g'-i')_0=0.873\pm 0.010$~mag and $(g'-i')_0=1.177\pm 0.007$~mag, and a separation between the blue and red GC populations at $(g'-i')_0=1.082$~mag. The color distributions for the GCSs of NGC~4709, NGC~4696, and the region between them are compared in Fig.~\ref{fig:comparison color distribution}. 

From these results, we can see that the red peak of the color distribution coincides within the error bars with that of the color distribution of the GCS of NGC~4709. This suggests that the properties of the GCs in the region are similar to those of NGC~4709. We then calculated the GCLF in this region assuming the same $\sigma$ derived for NGC~4709 (see Tab.~\ref{tab:LF}). The results of the Gaussian fits in the three bands are shown in Fig.~\ref{fig:LF between}, and the best-fit parameters are reported in Tab.~\ref{tab:region}. We found an error weighted mean distance between the three bands of $d_{\rm reg}= 34.69\pm 2.21$~Mpc, a total number of GC candidates of $N_{\rm GC}= 382$, and a distance modulus of $(m-M)=32.70\pm 0.14$. The distance is between that of NGC~4709 derived in this work and that for NGC~4696 derived in \citet{Federle}, suggesting the presence of a bridge of GCs connecting the two galaxies and that, despite the large difference in distance, NGC~4709 may have passed by NGC~4696 in the distant past. This is in good agreement with recent X-ray studies, which confirmed temperature excess in the region of Cen~45 and found the presence of shock-heated gas given by the interaction between Cen~45 and Cen~30 \citep{Veronica25}. A comparison between the positions of GC candidates and X-ray contours is shown in Fig.~\ref{fig:X-ray contours}.

\section{Conclusions}
\label{sec:conclusion} 

We have presented the analysis of the GCS of the elliptical galaxy NGC~4709. The measurements are based on deep {\em Magellan} 6.5m/MegaCam ($g'$, $r'$, $i'$) photometry. We obtained the following results: \begin{itemize}
    \item Applying a two color selection and limiting our study within a galactocentric radius $r<5\times r_{\rm eff}$, we identified a total of $556$ GC candidates. The GC system has a bimodal color distribution, with peaks at $(g'-i')_0=0.905\pm 0.009$~mag and $(g'-i')_0=1.170\pm 0.008$~mag, with the blue and red populations divided at $(g'-i')_0=1.044$~mag.
    \item The azimuthal distribution of the GC candidates shows peaks at the position angles PA$_1\sim 92^{\circ}$ and PA$_2\sim 293^{\circ}$. The position angle PA$_2$ coincides within the errors with the direction of NGC~4696, confirming that the interaction between these galaxies shaped not only the GCS of NGC~4696, as we found in \citet{Federle}, but also the GCS of NGC~4709. Moreover, these results are in agreement with the merger/interaction history outlined by previous X-ray works, such as a metallicity excess around NGC~4709 by \citet{Walker13}, and an asymmetric temperature variation in the X-ray gas with the hottest region coinciding with NGC~4709 by \citet{Churazov99}.
    \item After correcting the number of GCs for the completeness fraction and the background sources, we calculated the GCLF in the three filters and used it for our own estimation of the distance of NGC~4709, as well as the specific frequency. As can be seen from Tab.~\ref{tab:LF}, the distances estimated in the three filters are consistent within the errors. Moreover, our distance of $d=29.9\pm 2.1$~Mpc and distance modulus of $(m-M)=32.38\pm 0.15$~mag are consistent with the measurement of the distance of the Centaurus cluster obtained with both the GCLF and the surface brightness fluctuation method by \citet{mieske}. Our results show that, despite a small angular separation of $\sim 14.6$~arcmin ($\sim 181$~kpc at the distance of Centaurus), NGC~4709 is located in front of NGC~4696 and the other galaxies of the cluster. Moreover, the calculated distance places NGC~4709 outside of Centaurus $R_{500}$, defined as the radius inside which the density of the cluster is $500$ times the critical density of the Universe (e.g., \citealt{Evrard96}). Given that Centaurus has a radius of $R_{500}=0.826$~Mpc \citep{Piffaretti11}, the distance between NGC~4709 and NGC~4696 results of $\sim 10.2~R_{500}$. However, given the large error bars on the distances of NGC~4696 and NGC~4709, we note that the distance between the two galaxies could be lower than what we calculated in this work, which would be consistent with the redshifts for Cen~30 and Cen~45 estimated by previous X-ray studies (\citealt{Ota16}). Finally, we used the peak of the GCLF to estimate the total number of GCs in NGC~4709, assuming that the luminosity function is described by a symmetric Gaussian. We then used the number of clusters to estimate the specific frequency of the system, which resulted of $S_N=3.7\pm 0.5$. This value is well within the range for normal ellipticals and in good agreement with previous calculations (\citealt{mieske}).
    \item The GCs color distribution of the entire Field 1 (the blue rectangular region of $\sim 26.91\times 26.40$~arcmin$^2$ in Fig.~\ref{fig:Centaurus}) is bimodal, with blue and red peaks at colors $(g'-i')_0=0.856\pm 0.009$~mag and $(g'-i')_0=1.111\pm 0.010$~mag, and a color separation at $(g'-i')_0=1.000$~mag.
    \item The GCs density maps show overdensities of GCs corresponding to the positions of the galaxies NGC~4709, NGC~4696, NGC4706, WISEA~J124956.33-411536.8, and ESO~323-G~009. In particular, the red GC population of NGC~4709 appears more centrally concentrated with respect to the blue one, suggesting an in situ origin with fewer accreted GCs. On the other hand, NGC~4696 shows a larger blue GC population, suggesting a more turbulent past, characterized by multiple mergers and accretion events. Over the entire Field 1, the blue GC population looks more uniformly distributed with respect to the red one. 
    \item We analyzed the color distributions for the GCSs of the galaxies corresponding to the overdensities identified in the density maps. We found that the GCSs of NGC~4696 and ESO~323-G~009 show a bimodal color distribution, whereas NGC~4706 has a unimodal color distribution. The number of GC candidates in WISEA~J124956.33-411536.8 is too low to obtain a reliable color distribution. 
    \item The color distribution of GC candidates located in a rectangular region between NGC~4709 and NGC~4696 is bimodal, with peaks at $(g'-i')_0=0.873\pm 0.010$~mag and $(g'-i')_0=1.177\pm 0.007$~mag, with a separation between the blue and red populations at $(g'-i')_0=1.082$~mag. It is worth noticing that the red peak of the color distribution in this region coincides within the errors with that of the GCS of NGC~4709, suggesting that their GCs have similar properties. We then calculated the GCLF in this region using the same $\sigma$ as for the GCLF of NGC~4709, and we used it to estimate the distance of the GC candidates, which resulted of $d\sim 34.69\pm 2.21$~Mpc. This value is between the distance of NGC~4709 estimated in this work and that of NGC~4696 that we obtained in \citet{Federle}, suggesting the presence of a bridge of GCs connecting the two galaxies. 
\end{itemize}

As a final remark, given our findings and those in literature X-ray studies (\citealt{Churazov99}; \citealt{Walker13}; \citealt{Veronica25}), we propose that NGC~4709 interacted with NGC~4696 in the past in a high-speed flyby, in which the two galaxies had a fast close encounter. During the encounter, NGC~4709 retained its original gas but was stripped of some GCs that formed a bridge between the two galaxies. NGC~4709 was afterward captured by the Cen~45 group, and it is now going back toward Cen~30. In a future work, we will compare our results with E-MOSAICS simulations to confirm this scenario.

\section{Data availability}

The full catalog of GC candidates (Table~\ref{tab:appendix}) is only available in electronic form at the CDS via anonymous ftp to \url{cdsarc.u-strasbg.fr} (130.79.128.5) or via \url{http://cdsweb.u-strasbg.fr/cgi-bin/qcat?J/A+A/}.

\begin{acknowledgements}
S.~F. and M.~G. gratefully acknowledge support by ANID through Fondecyt Regular $1240755$. \\
This research has made use of the NASA/IPAC Extragalactic Database (NED), which is funded by the National Aeronautics and Space Administration and operated by the California Institute of Technology.

\end{acknowledgements}

\clearpage

\onecolumn
\begin{appendix}

\begin{sidewaystable}[h!]
\section{Catalog of globular cluster candidates}
     \caption{Properties of the GC candidates in Field 1 of the Centaurus cluster.}
    \centering
   \begin{adjustbox}{width=\textheight}
   \begin{tabular}{c c c c c c c c c c c c c c c c c c}
    \hline
     RA & DEC & {\footnotesize{CLASS$_{g'}$}} & {\footnotesize{CLASS$_{r'}$}} & {\footnotesize{CLASS$_{i'}$}} & $g'_0$ & $r'_0$ & $i'_0$ & $r_{\rm NGC~4709}$ & $r_{\rm NGC~4706}$ & $r_{\rm NGC~4696}$ & $r_{\rm WISEA}$ & $r_{\rm ESO~323}$ & PA$_{\rm NGC~4709}$ & PA$_{\rm NGC~4706}$ & PA$_{\rm NGC~4696}$ & PA$_{\rm WISEA}$ & PA$_{\rm ESO~323}$ \\
    (hh:mm:ss) & ($^{\circ}$:':'') & & & & (mag) & (mag) & (mag) & (arcmin) & (arcmin) & (arcmin) & (arcmin) & (arcmin) & ($^{\circ}$) &  ($^{\circ}$) & ($^{\circ}$) & ($^{\circ}$) & ($^{\circ}$) \\
    \hline
    12:49:12.17 & -41:39:50.11 & 0.686 & 0.011 & 0.610 & $24.416\pm 0.070$ & $23.813\pm 0.034$ & $23.559\pm 0.057$ & 19.492 & 24.378 & 21.616 & 25.596 & 22.028 & 209.776 & 198.805 & 168.443 & 198.831 & 230.499\\
  12:49:46.33 & -41:39:49.52 & 0.234 & 0.979 & 0.968 & $23.983\pm 0.029$ & $23.348\pm 0.018$ & $23.113\pm 0.028$ & 17.224 & 23.110 & 23.721 & 24.284 & 17.566 & 191.047 & 183.668 & 153.152 & 184.445 & 217.187\\
  12:50:06.11 & -41:39:46.28 & 0.549 & 0.893 & 0.607 & $25.261\pm 0.089$ & $24.513\pm 0.046$ & $24.356\pm 0.084$ & 16.858 & 23.118 & 25.560 & 24.227 & 15.566 & 178.662 & 174.499 & 145.688 & 175.711 & 206.414\\
  12:49:17.20 & -41:39:46.21 & 0.503 & 0.070 & 0.585 & $25.298\pm 0.091$ & $24.676\pm 0.055$ & $24.243\pm 0.076$ & 18.984 & 24.028 & 21.759 & 25.244 & 21.267 & 207.412 & 196.734 & 165.981 & 196.860 & 229.026\\
  12:48:50.75 & -41:39:44.34 & 0.965 & 0.976 & 0.975 & $22.690\pm 0.015$ & $22.155\pm 0.010$ & $22.000\pm 0.012$ & 21.690 & 25.867 & 21.091 & 27.074 & 25.195 & 219.105 & 207.287 & 179.107 & 206.931 & 236.449\\
  12:48:38.52 & -41:39:41.79 & 0.680 & 0.955 & 0.784 & $24.890\pm 0.062$ & $24.208\pm 0.034$ & $23.954\pm 0.057$ & 23.172 & 26.960 & 21.142 & 28.151 & 27.110 & 223.547 & 211.639 & 185.306 & 211.111 & 239.180\\
  12:50:30.40 & -41:39:45.54 & 0.053 & 0.966 & 0.786 & $22.060\pm 0.007$ & $21.442\pm 0.005$ & $21.222\pm 0.006$ & 17.555 & 23.978 & 28.363 & 24.974 & 14.138 & 163.690 & 163.642 & 138.088 & 165.275 & 189.738\\
  12:50:45.35 & -41:39:39.97 & 0.700 & 0.013 & 0.605 & $25.308\pm 0.069$ & $24.897\pm 0.054$ & $24.463\pm 0.076$ & 18.456 & 24.830 & 30.240 & 25.746 & 13.855 & 155.267 & 157.393 & 134.039 & 159.207 & 178.355\\
  12:50:22.89 & -41:39:40.29 & 0.644 & 0.846 & 0.863 & $25.039\pm 0.057$ & $24.306\pm 0.032$ & $24.206\pm 0.060$ & 17.125 & 23.533 & 27.376 & 24.567 & 14.355 & 168.114 & 166.860 & 140.148 & 168.387 & 195.322\\
  12:48:54.33 & -41:39:39.14 & 0.730 & 0.839 & 0.919 & $25.041\pm 0.056$ & $24.428\pm 0.036$ & $24.159\pm 0.057$ & 21.204 & 25.488 & 21.024 & 26.698 & 24.591 & 217.853 & 206.041 & 177.279 & 205.737 & 235.759\\
  12:50:18.55 & -41:39:40.04 & 0.949 & 0.961 & 0.963 & $23.841\pm 0.026$ & $23.039\pm 0.013$ & $22.635\pm 0.019$ & 16.971 & 23.356 & 26.860 & 24.411 & 14.584 & 170.787 & 168.792 & 141.467 & 170.246 & 198.395\\
  \hline
    \end{tabular}
    \end{adjustbox}
    \tablefoot{The columns show the coordinates of the sources (RA and DEC), the stellarity index (CLASS$_{g'}$, CLASS$_{r'}$, and CLASS$_{i'}$), the magnitudes and errors in each band ($g'_0$, $r'_0$, and $i'_0$), the distance from the center of each galaxy considered ($r_{\rm NGC~4709}$, $r_{\rm NGC~4706}$, $r_{\rm NGC~4696}$, $r_{\rm WISEA}$, and $r_{\rm ESO~323}$), and the position angle of the GC candidates with respect to each galaxy (PA$_{\rm NGC~4709}$, PA$_{\rm NGC~4706}$, PA$_{\rm NGC~4696}$, PA$_{\rm WISEA}$, and PA$_{\rm ESO~323}$). Here, for simplicity, we abbreviated WISEA~J124956.33-411536.8 in WISEA and ESO~323-G~009 as ESO~323.}
    \label{tab:appendix}
\end{sidewaystable}

\clearpage

\section{Parameters of the galaxies and of their GCSs}

\begin{table*}[h!]
    \caption{Parameters of the galaxies NGC~4696, NGC~4706, WISEA~J124956.33-411536.8, and ESO~323-G~009.}
    \centering
    \resizebox{\textwidth}{!}{\begin{tabular}{c c c c c c c c c}
    \hline
    Name & RA & DEC & Type & $d$ & PA & $r_{\rm eff}$ & $e$ & $e_{\rm eff}$ \\
     & (J2000) & (J2000) & (RC3) & (arcmin) & ($^{\circ}$) & (arcmin) & & \\
     \hline
    NGC~4696 & $12^{\rm h}48^{\rm m}49^{\rm s}.80$ & $-41^{\circ}18'39''.00$ & cD pec & $14.64$ & $286.85$ & $0.934$ & $0.135$ & $0.215\pm 0.001$ \\
    NGC~4706 & $12^{\rm h}49^{\rm m}54^{\rm s}.18$ & $-41^{\circ}16'45''.71$ & SAB0${\text{\textasciicircum}}$0(s) edge-on & $6.42$ & $343.5$ & $0.223$ & $0.46$ & $0.317\pm 0.002$ \\
    WISEA~J124956.33-411536.8 & $12^{\rm h}49^{\rm m}56^{\rm s}.31$ & $-41^{\circ}15'37''.10$ & dE & $7.44$ & $340.02$ & $0.152$ & - & $0.244\pm 0.004$ \\
    ESO~323-G~009 & $12^{\rm h}50^{\rm m}42^{\rm s}.99$ & $-41^{\circ}25'49''.08$ & S0 edge-on & $7.89$ & $111.63$ & $0.314$ & $0.42$ & $0.502\pm 0.008$ \\ 
    \hline
    \end{tabular}}
    \tablefoot{The columns show the name, coordinates of the center (RA and DEC), type, and ellipticity reported in NED. The remaining columns show the effective radius ($r_{\rm eff}$), ellipticity at the effective radius ($e_{\rm eff}$), distance ($d$), and position angle with respect to NGC~4709 (PA) calculated in this work.}
    \label{tab:galaxies parameters}
\end{table*}

\begin{table*}[h!]
    
    \caption{Shape of the GCSs of the galaxies in Tab.~\ref{tab:galaxies parameters}.}
    \centering
    \resizebox{\textwidth}{!}{\begin{tabular}{c c c c c c c c}
    \hline
    Name & $N_{\rm GC}$ & RA & DEC & $a$ & $b$ & $e$ & $e_{\rm eff}$ \\
     & & (J2000) & (J2000) & (arcmin) & (arcmin) & & \\
    \hline
    NGC~4696 & $1289$ & $12^{\rm h}49^{\rm m}03^{\rm s}.13^{+0.07}_{-0.05}$ & $-41^{\circ}20'12''.08^{+3.45}_{-4.65}$ & $5.76^{+0.02}_{-0.01}$ & $4.94^{+0.06}_{-0.05}$ & $0.143^{+0.009}_{-0.011}$ & $0.215\pm 0.001$ \\
    NGC~4706 & $46$ & $12^{\rm h}49^{\rm m}54^{\rm s}.54^{+0.16}_{-0.09}$ & $-41^{\circ}16'47''.36^{+2.05}_{-4.20}$ & $0.94^{+0.04}_{-0.05}$ & $0.56\pm 0.04$ & $0.405^{+0.051}_{-0.043}$ & $0.317\pm 0.002$ \\
     WISEA~J124956.33-411536.8 & $11$ & $12^{\rm h}49^{\rm m}56^{\rm s}.60^{+0.52}_{-0.52}$ & $-41^{\circ}15'42''.24^{+0.80}_{-0.60}$ & $0.50^{+0.07}_{-0.09}$ & $0.37\pm 0.02$ & $0.240^{+0.090}_{-0.080}$ & $0.244\pm 0.004$ \\
    ESO~323-G~009 & $64$ & $12^{\rm h}50^{\rm m}43^{\rm s}.43^{+0.11}_{-0.05}$ & $-41^{\circ}25'45''.16^{+1.25}_{-3.00}$ & $1.32^{+0.02}_{-0.01}$ & $0.91\pm 0.03$ & $0.312^{+0.026}_{-0.019}$ & $0.502\pm 0.008$ \\
    \hline
    \end{tabular}}

    \tablefoot{The columns show the name of the galaxy, the number of GC candidates ($N_{\rm GC}$), the coordinates of the center of the GCS (RA and DEC), the major and minor axes ($a$ and $b$), and the ellipticity ($e$). For comparison, the last column shows the galaxy's ellipticity at the effective radius.}
    \label{tab:shape GCSs}
\end{table*}

\section{Color-magnitude diagram}

\begin{SCfigure}[1][h]
    \centering
    \includegraphics[width=0.5\textwidth]{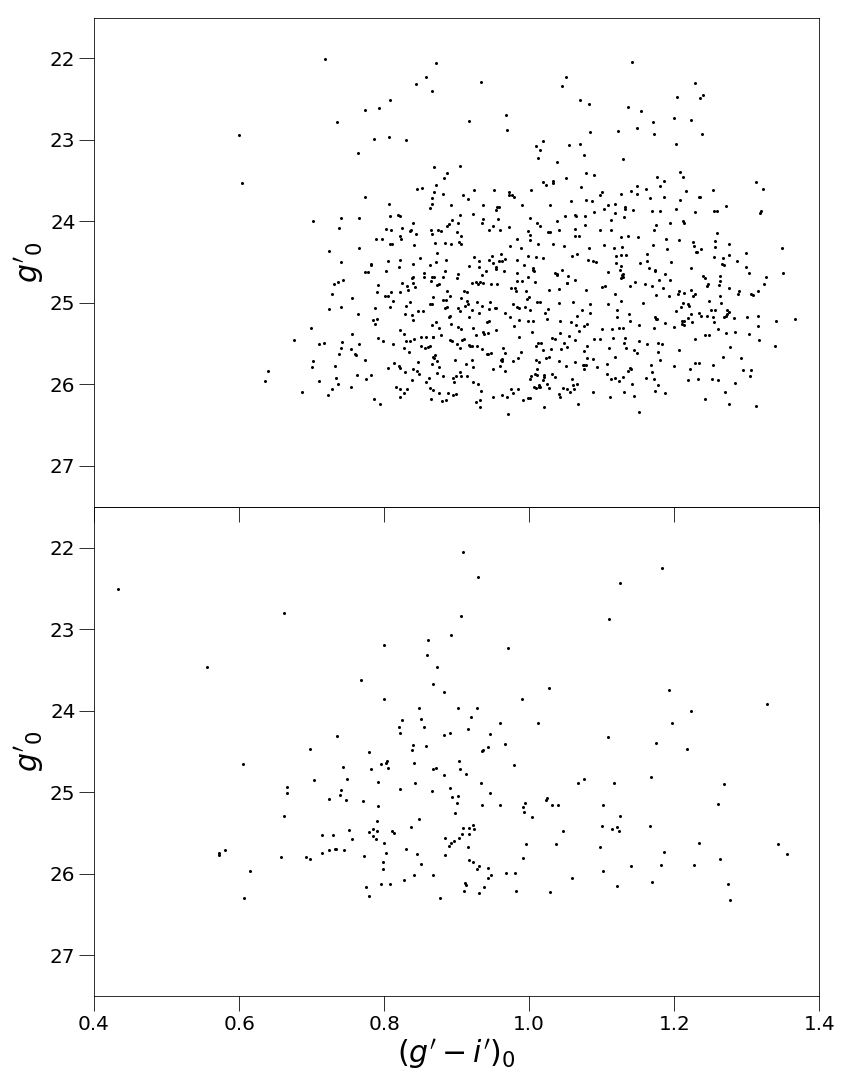}
    \caption{Comparison between the color-magnitude diagram of the GC candidates of NGC~4709 and that of the region between NGC~4709 and NGC~4696. {\em Top panel}: Color-magnitude diagram of the $785$ GC candidates within a galactocentric distance of $5\times r_{\rm eff}$ from the center of NGC~4709. {\em Bottom panel}: Color-magnitude diagram of the $202$ GC candidates in the $\sim 20.89$~arcmin$^2$ region between NGC~4709 and NGC~4696.}
    \label{fig:cmd bet}
\end{SCfigure}

\clearpage
\onecolumn
\section{Color distributions}

\begin{figure*}[h!]
    \centering
    \includegraphics[scale=0.27]{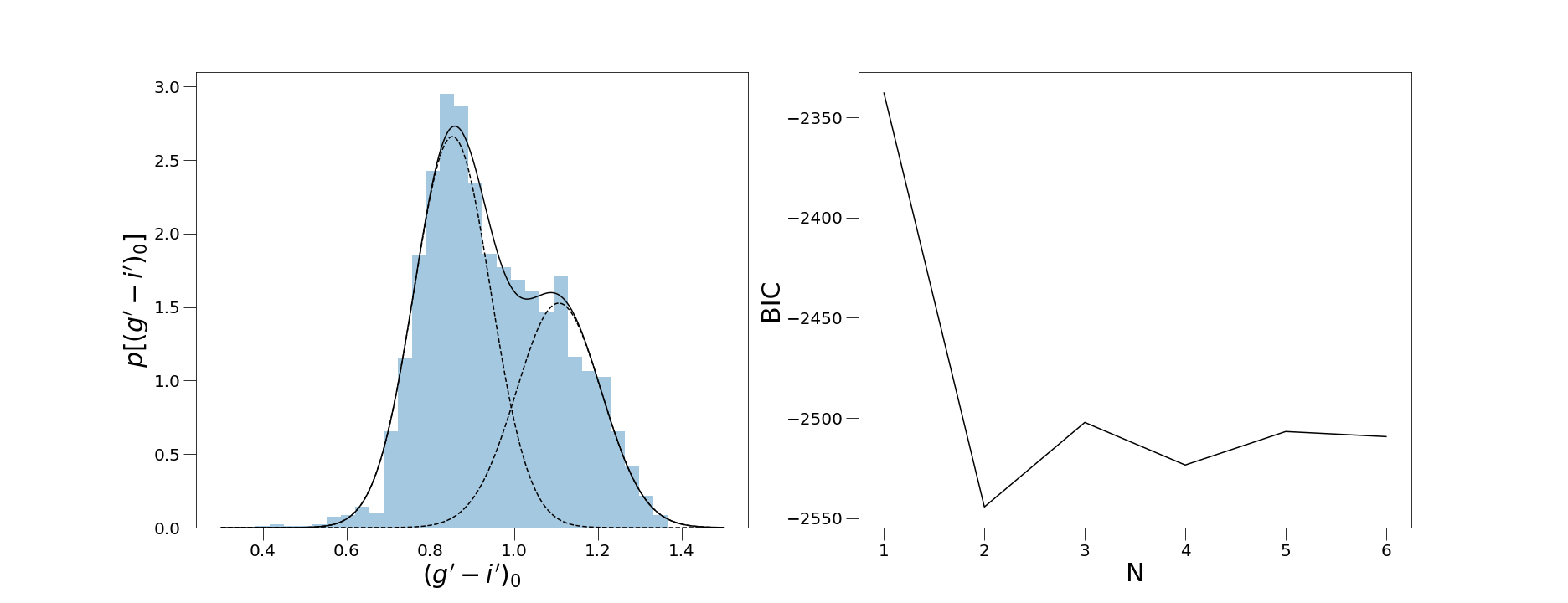}
    \caption{Bayesian Information Criterion in Field 1. {\em Left Panel}: probability density function vs the $(g'-i')_0$ color. The black line represents the best-fit model, and the dashed lines represent the Gaussians described by the model. The color distribution is divided in bins according to the Freedman-Diaconis rule (\citealt{Freedman81}). {\em Right panel}: values of the BIC as a function of the number of components in the model.}
    \label{fig:total color}
\end{figure*}

\begin{figure*}[h!]
    \centering
    \includegraphics[scale=0.25]{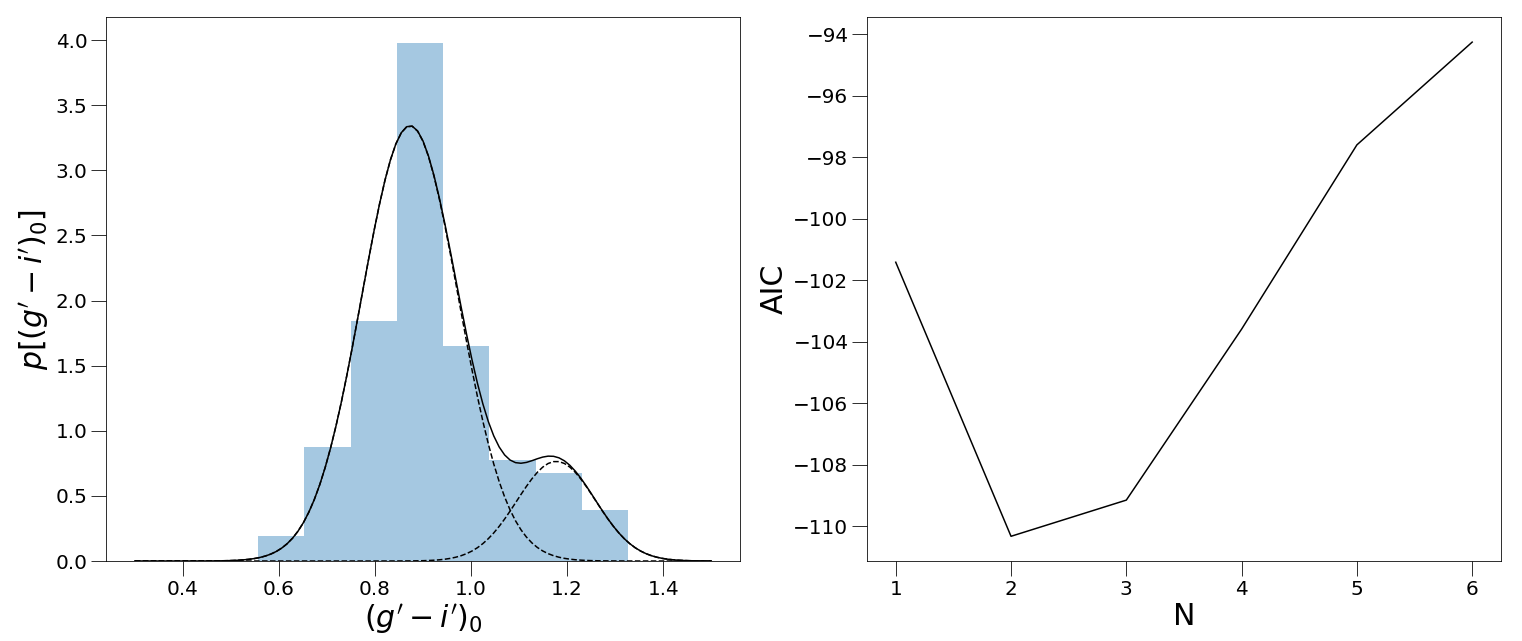}
    \caption{Akaike Information Criterion test on the GC candidates in the region between NGC~4709 and NGC~4696. {\em Left panel}: probability density function vs $(g'-i')_0$ color. The black line represents the best-fit model, and the dashed lines represent the two Gaussians described by the model. The color distribution is divided in bins according to the Freedman-Diaconis rule (\citealt{Freedman81}). {\em Right panel}: values of the Akaike Information Criterion (AIC) as a function of the number of components in the model.}
    \label{fig:AIC}
\end{figure*}

\begin{table*}[h!]
    \caption{Parameters of the color distributions for the galaxies in Tab.\ref{tab:galaxies parameters}, and for the region between NGC~4709 and NGC~4696.}
    \centering
    \begin{tabular}{c c c c c c c}
    \hline
    Name & $N_{\rm GC}$ & $\mu_{\rm red}$ & $\mu_{\rm blue}$ & $\mu_{\rm sep}$ & $N_{\rm red}$ & $N_{\rm blue}$  \\
      & & (mag) & (mag) & (mag) & & \\
     \hline
    NGC~4696 & $1289$ & $1.094\pm 0.010$ & $0.850\pm 0.008$ & $0.987$ & $471$ & $815$ \\
    NGC~4706 & $49$ & - & - & $0.929\pm 0.028$ & - & - \\
    WISEA~J124956.33-411536.8 & $11$ & - & - & - & - & - \\ 
    ESO~323-G~009 & $64$ & $1.041\pm 0.006$ & $0.831\pm 0.002$ & $0.917$ & $29$ & $35$ \\
    Region & $107$ & $1.177\pm 0.007$ & $0.873\pm 0.010$ & $1.082$ & $17$ & $90$ \\
    \hline
    \end{tabular}
    \tablefoot{The columns show the name of the galaxy, the total number of GC candidates ($N_{\rm GC}$), the $(g'-i')_0$ color of the blue and red peaks ($\mu_{\rm red}$ and $\mu_{\rm blue}$), the color separation between the blue and red populations ($\mu_{\rm sep}$), the number of red and blue GC candidates ($N_{\rm red}$ and $N_{\rm blue}$).}
    \label{tab:color dis galaxies}
\end{table*}
\clearpage
\twocolumn
\begin{figure}
    \centering
    \includegraphics[scale=0.29]{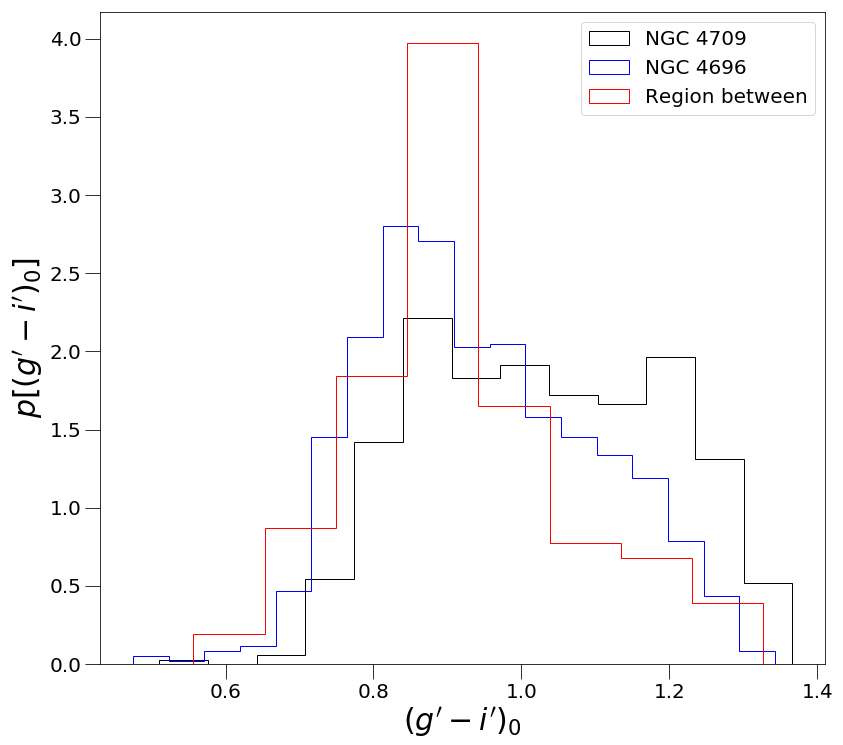}
    \caption{Comparison of the color distributions. The plot shows the probability density function ($p[(g'-i')_0]$) as a function of the $(g'-i')_0$ color for NGC~4709 (black line), NGC~4696 (blue line) and the rectangular region between them (red line) shown in red in Fig.\ref{fig:positions}. The three color distributions are normalized so that the area of each histogram is equal to $1$.}
    \label{fig:comparison color distribution}
\end{figure}

\columnbreak

\section{Luminosity function of the region between NGC~4709 and NGC~4696}

\begin{figure}[H]
    \centering
    \includegraphics[scale=0.25]{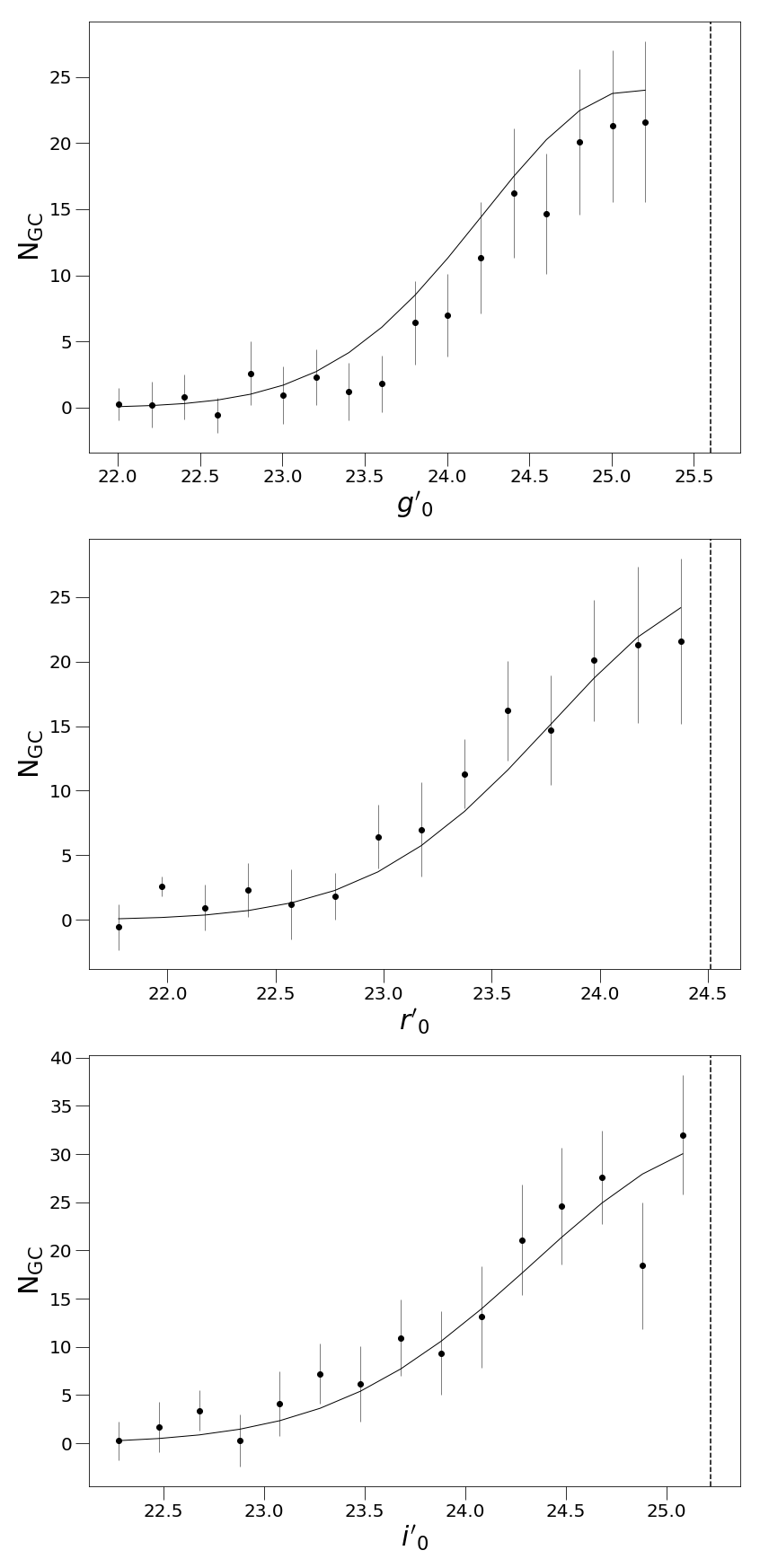}
    \caption{GCLF of the region between NGC~4709 and NGC~4696. {\em Top, Middle, and Bottom panels}: GCLF in the $g'_0$, $r'_0$, and $i'_0$ bands, where the black dots represent the number of clusters corrected for the completeness fraction and for the background level divided in bins of $0.2$~mag, and the solid line represents the best-fit Gaussian. The dashed lines show the magnitudes at which we have a $67\%$ completeness.}
    \label{fig:LF between}
\end{figure}
\onecolumn
\begin{table*}[h!]
    \caption{Best-fit parameters for the GCLF of the region between NGC~4709 and NGC~4696.}
    \centering
    \begin{tabular}{c c c c c c}
    \hline
    Band & mag & $A$ & $\mu$ & $d$ & $(m-M)$  \\
     & (mag) & & (mag) & (Mpc) & (mag) \\
     \hline
    $g'_0$ & $25.34$ & $6.46\pm 0.13$ & $23.6\pm 0.15$ & $35.36\pm 3.55$ & $32.74\pm 0.22$ \\
    $r'_0$ & $24.51$ & $25.32^{+0.29}_{-0.30}$ & $24.85\pm 0.02$ & $30.39\pm 3.22$ & $32.41\pm 0.23$\\
    $i'_0$ & $25.22$ & $31.06^{+0.94}_{-0.89}$ & $25.54\pm 0.03$ & $46.91\pm 5.81$ & $33.36\pm 0.27$ \\
    \hline
    \end{tabular}
    \tablefoot{The columns show the band, the magnitude up to which the LF was fitted (mag), the amplitude ($A$), the mean ($\mu$), and the calculated distance ($d$) and distance modulus ($m-M$) for the region between NGC~4709 and NGC~4696.}
    \label{tab:region}
\end{table*}

\section{X-ray contours}

\begin{figure}[h!]
    \centering
    \includegraphics[scale=0.7]{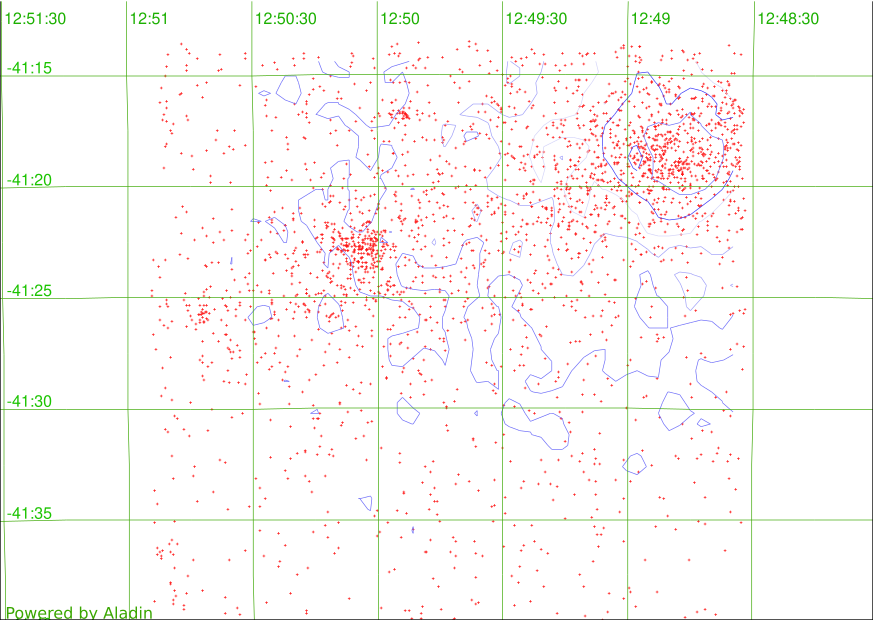}
    \caption{X-ray contours. The plot shows the positions of the GC candidates obtained in this work (red dots) compared to the X-ray contours obtained from eROSITA (blue lines). The X-ray contours were obtained from data by \citet{Veronica25}. The plot was made using Aladin (\citealt{Bonnarel2000}; \citealt{Boch14}).}
    \label{fig:X-ray contours}
\end{figure}


\end{appendix}


\begin{thebibliography}{}
\bibitem[Akaike (1974)]{Akaike}
Akaike, H. \ 1974, ITAC, 19, 716
\bibitem[Akhil, Kartha \& Blesson (2024)]{Akhil24}
Akhil, K.~R., Kartha, S.~S., \& Mathew, B. \ 2024, MNRAS, 530, 2907
\bibitem[Alamo-Martinez et al.(2013)]{Alamo-Martinez13}
Alamo-Martinez, K.~A., Blakeslee, J.~P., C{\^o}té, P., et al. \ 2013, ApJ, 775, 20
\bibitem[Astropy Collaboration et al.(2013)]{astropy1}
Astropy Collaboration, Robitaille, T.~P., Tollerud, E.~J., Greenfield, P., et al. \ 2013, A\&A, 558, A33
\bibitem[Astropy Collaboration et al.(2018)]{astropy2}
Astropy Collaboration, Price-Whelan, A.~M., Sip{\H{o}}cz, B.~M., et al. \ 2018, AJ, 156, 123
\bibitem[Beasley et al.(2018)]{Beasley18}
Beasley, M.~A., Trujillo, I., Leaman, R., et al. \ 2018, Nat, 555, 483
\bibitem[Beasley (2020)]{Beasley20}
Beasley, M.~A. \ 2020, rfma.book, 245. doi:10.1007/978-3-030-38509-5\_9
\bibitem[Bertin et al.(2002)]{swarp}
Bertin, E., Mellier, Y., Radovich, M., et al. \ 2002, ASPC, 281, 228
\bibitem[Bertin \& Arnouts (1996)]{sex}
Bertin, E. \& Arnouts, S. \ 1996, A\&AS, 117, 393
\bibitem[Binney \& Wong (2017)]{Binney17}
Binney, J. \& Wong, L.~K. \ 2017, MNRAS, 467, 2446
\bibitem[Blom et al.(2014)]{Blom14}
Blom, C., Forbes, D.~A., Foster, C., et al. \ 2014, MNRAS, 439, 2420
\bibitem[Boch \& Fernique (2014)]{Boch14}
Boch, T., \& Fernique, P. / 2014, ASPC, 485, 277
\bibitem[Bonnarel et al.(2000)]{Bonnarel2000}
Bonnarel, F., Fernique, P., Bienaymé, O., et al. / 2000, A\&AS, 143, 33
\bibitem[Brodie \& Strader (2006)]{Brodie06}
Brodie, J.~P. \& Strader, J. \ 2006, ARA\&A, 44, 193
\bibitem[Chiboucas \& Mateo (2007)]{Chiboucas07}
Chiboucas, K. \& Mateo, M. \ 2007, ApJS, 170, 95
\bibitem[Churazov et al.(1999)]{Churazov99}
Churazov, E., Gilfanov, M., Forman, W., et al. \ 1999, ApJ, 520, 105
\bibitem[Ciambur (2015)]{ciambur}
Ciambur, B.~C. \ 2015, ApJ, 810, 120
\bibitem[C{\^o}té, Marzke \& West (1998)]{Cote98}
C{\^o}té, P., Marzke, R.~O. \& West, M.~J. \ 1998, ApJ, 501, 554
\bibitem[de Vaucouleurs et al.(1991)]{deV}
de Vaucouleurs, G., de Vaucouleurs, A., Corwin, H.~G., Jr., et al. \ 1991, Third Reference Catalogue of Bright Galaxies (New York, NY: Springer)
\bibitem[Dias et al.(2022)]{Dias22}
Dias, B., Palma, T., Minniti, D., et al. \ 2022, A\&A, 657, A67
\bibitem[Dolfi et al.(2021)]{Dolfi21}
Dolfi, A., Forbes, D.~A., Couch, W.~J., et al. \ 2021, MNRAS, 504, 4923
\bibitem[Escudero et al.(2018)]{Escudero18}
Escudero, C.~G., Faifer, F.~R., Smith Castelli, A.~V., et al. \ 2018, MNRAS, 474, 4302
\bibitem[Escudero et al.(2022)]{Escudero22}
Escudero, C.~G., Cortesi, A., Faifer, F.~R., et al. \ 2022, MNRAS, 511, 393
\bibitem[Evrard, Metzler \& Navarro (1996)]{Evrard96}
Evrard, A.~E., Metzler, C.~A. \& Navarro, J.~F. \ 1996, ApJ, 469, 494
\bibitem[Faifer et al.(2011)]{Faifer11}
Faifer, F.~R., Forte, J.~C., Norris, M.~A., et al. \ 2011, MNRAS, 416, 155
\bibitem[Faifer et al.(2017)]{Faifer17}
Faifer, F.~R., Escudero, C.~G., Scalia, M.~C., et al. \ 2017, A\&A, 599, L8
\bibitem[Federle et al.(2024)]{Federle}
Federle, S., G{\'o}mez, M., Mieske, S., et al. \ 2024, A\&A, 689, A342
\bibitem[Freedman \& Diaconis (1981)]{Freedman81}
Freedman, D. \& Diaconis, P. \ 1981, "On the histogram as density estimator: L2 theory", Probability Theory and Related Fields 57 (4): 453-476
\bibitem[Garro et al.(2021)]{Garro21}
Garro, E.~R., Minniti, D., G{\'o}mez, M., et al. \ 2021, A\&A, 649, A86
\bibitem[Garro, Minniti \& Fern{\'a}ndez-Trincado (2024)]{Garro24}
Garro, E.~R., Minniti, D. \& Fern{\'a}ndez-Trincado, J.~G. \ 2024, A\&A, 687, A214
\bibitem[Georgiev et al.(2010)]{Georgiev10}
Georgiev, I.~Y., Puzia, T.~H., Goudfrooij, P., et al. \ 2010, MNRAS, 406, 1967
\bibitem[Gnedin (2003)]{Gnedin03}
Gnedin, O.~Y. \ 2003, ApJ, 582, 141
\bibitem[Harris \& van den Bergh (1981)]{Harris81}
Harris, W.~E. \& van den Bergh, S. \ 1981, AJ, 86, 1627
\bibitem[Harris (1991)]{Harris91}
Harris, W.~E. \ 1991, ARA\&A, 29, 543
\bibitem[Harris (2010)]{Harris10}
Harris, W.~E. \ 2010, arXiv e-prints, arXiv:1012.3224. doi:10.48550/arXiv.1012.3224
\bibitem[Hughes et al.(2023)]{Hughes23}
Hughes, A.~K., Sand, D.~J., Seth, A., et al. \ 2023, ApJ, 947, 34
\bibitem[Ishchenko et al.(2023)]{Ishchenko23}
Ishchenko, M., Sobolenko, M., Berczik, P., et al. \ 2023, A\&A, 673, A152
\bibitem[Janssens et al.(2024)]{Janssens24}
Janssens, S.~R., Forbes, D.~A., Romanowsky, A.~J., et al. \ 2024, MNRAS, 534, 783
\bibitem[Jedrzejewski (1987)]{jed}
Jedrzejewski, R.~I. \ 1987, MNRAS, 226, 747
\bibitem[Jord{\'a}n et al.(2007)]{Jordan07}
Jord{\'a}n, A., McLaughlin, D.~E., C{$\hat{\rm o}$}t{\'e}, P., et al. \ 2007, ApJS, 171, 101
\bibitem[Kaviraj et al.(2012)]{Kaviraj12}
Kaviraj, S., Crockett, R.~M., Whitmore, B.~C., et al. \ 2012, MNRAS, 422, L96
\bibitem[Lahén et al.(2018)]{Lahén18}
Lahén, N., Johansson, P.~H., Rantala, A., et al. \ 2018, MNRAS, 475, 3934
\bibitem[Lauberts \& Valentijn (1989)]{Lauberts89}
Lauberts, A. \& Valentijn, E.~A. \ 1989, spce.book
\bibitem[Lomel{\'\i}-N{\'u}{\~n}ez et al.(2025)]{Lomeli25}
Lomel{\'\i}-N{\'u}{\~n}ez, L., Cortesi, A., Smith Castelli, A.~V., et al. \ 2025, AJ, 169, 263
\bibitem[Lonare et al.(2025)]{Lonare25}
Lonare, P., Cantiello, M., Mirabile, M., et al. \ 2025, A\&A, 694, A231
\bibitem[Lucey et al.(1986)]{Lucey86}
Lucey, J.~R., Currie, M.~J. \& Dickens, R.~J. \ 1986, MNRAS, 221, 453
\bibitem[McLeod et al.(2006)]{Megacam}
McLeod, B., Geary, J., Ordway, M., et al. \ 2006, ASSL, 336, 337
\bibitem[McLeod et al.(2015)]{Megacam15}
McLeod, B., Geary, J., Conroy, M. \ 2015, PASP, 127, 366
\bibitem[Mieske \& Hilker (2003)]{Mieske03}
Mieske, S. \& Hilker, M. \ 2003, A\&A, 410, 445
\bibitem[Mieske, Hilker, \& Infante (2005)]{mieske}
Mieske, S., Hilker, M. \& Infante, L. \ 2005, A\&A, 438, 103 
\bibitem[Mihos (2004)]{Mihos04}
Mihos, J.~C. \ 2004, cgpc.symp, 277
\bibitem[Minniti et al.(2010)]{Minniti10}
Minniti, D., Lucas, P.~W., Emerson, J.~P., et al. \ 2010, NewA, 15, 433. doi:10.1016/j.newast.2009.12.002
\bibitem[Minniti (2018)]{Minniti18}
Minniti, D. \ 2018, ASSP, 51, 63
\bibitem[Minniti, Palma \& Clariá (2021)]{Minniti21}
Minniti, D., Palma, T. \& Clariá, J.~J. \ 2021, BAAA, 62, 107
\bibitem[Misgeld, Hilker \& Mieske (2009)]{Misgeld09}
Misgeld, I., Hilker, M., \& Mieske, S. \ 2009, A\&A, 496, 683
\bibitem[Nidever et al.(2021)]{nsc}
Nidever, D.~L., Dey, A., Fasbender, K., et al. \ 2020, ApJ, 161, 192
\bibitem[Obasi et al.(2023)]{Obasi23}
Obasi, C., G{\'o}mez, M., Minniti, D., et al. \ 2023, A\&A, 670, A18
\bibitem[Oldham \& Auger (2016)]{Oldham16}
Oldham, L.~J. \& Auger, M~W. \ 2016, MNRAS, 455, 820
\bibitem[Ota \& Yoshida (2016)]{Ota16}
Ota, N. \& Yoshida, H. \ 2016, PASJ, 68, S19
\bibitem[Pedregosa et al.(2011)]{Pedregosa11}
Pedregosa, F., Varoquaux, G., Gramfort, A., et al. \ 2011, JMLR, 12, 2825
\bibitem[Peng et al.(2008)]{Peng08}
Peng, E.~W., Jord{\'a}n, A., C{$\hat{\rm o}$}t{\'e}, P., et al. \ 2008, ApJ, 681, 197
\bibitem[Pfeffer et al.(2018)]{Pfeffer18}
Pfeffer, J., Krujissen, J.~M.~D., Crain, R.~A., et al. \ 2018, MNRAS, 475, 4309 
\bibitem[Piffaretti et al.(2011)]{Piffaretti11}
Piffaretti, R., Arnaud, M., Pratt, G.~W., et al. \ 2011, A\&A, 534, A109
\bibitem[Proctor et al.(2009)]{Proctor09}
Proctor, R.~N., Forbes, D.~A., Romanowsky, A.~J., et al. \ 2009, MNRAS, 368, 91
\bibitem[Puzia et al.(2002)]{Puzia02}
Puzia, T.~H., Zepf, S.~E., Kissler-Patig, M., et al. \ 2002, A\&A, 391, 453
\bibitem[Racine (1968)]{Racine68}
Racine, R. \ 1968, JRASC, 62, 367
\bibitem[Reina-Campos et al.(2022)]{Reina-Campos22}
Reina-Campos, M., Trujillo-Gomez, S., Deason, A.~J., et al. \ 2022, MNRAS, 513, 3925
\bibitem[Reina-Campos et al.(2023)]{Reina-Campos23}
Reina-Campos, M., Trujillo-Gomez, S., Pfeffer, J.~L., et al. \ 2023, MNRAS, 521, 6368
\bibitem[Rejkuba (2012)]{Rejkuba12}
Rejkuba, M., \ 2012, Ap\&SS, 341, 195
\bibitem[Renaud, Agertz \& Gieles (2017)]{Renaud17}
Renaud, F., Agertz, O. \& Gieles, M. \ 2017, MNRAS, 465, 3622
\bibitem[Richtler et al.(2014)]{Richtler14}
Richtler, T., Hilker, M., Kumar, B., et al. \ 2014, A\&A, 569, A41
\bibitem[Robin et al.(2003)]{Robin03}
Robin, A.~C., Reylé, C., Derrière, S., \& Picaud, S. \ 2003, A\&A, 409, 523
\bibitem[Schlafly \& Finkbeiner (2011)]{SF2011}
Schlafly, E.~F. \& Finkbeiner, D.~P. \ 2011, ApJ, 737, 103  
\bibitem[Schwarz (1978)]{Schwarz}
Schwarz, G. \ 1978, "Estimating the Dimension of a Model" Ann. Statist. 6 (2) 461-464
\bibitem[Sesto, Faifer \& Forte (2016)]{Sesto16}
Sesto, L.~A., Faifer, F.~R. \& Forte, J.~C. \ 2016, MNRAS, 461, 4260
\bibitem[Stein, Jerjen \& Federspiel (1997)]{Stein97}
Stein, P., Jerjen, H. \& Federspiel, M. \ 1997, A\&A, 327, 952
\bibitem[Stetson (1987)]{Stetson87}
Stetson, P.~B. \ 1987, PASP, 99, 191
\bibitem[Taylor et al.(2017)]{Taylor17}
Taylor, M.~A., Puzia, T.~H., Mu{\~n}oz, R.~P., et al. \ 2017, MNRAS, 469, 3444
\bibitem[Tully et al.(2013)]{Tully13}
Tully, R.~B., Courtois, H.~M., Dolphin, A.~E., et al. \ 2013, AJ, 146, 86
\bibitem[Urbano et al.(2024)]{Urbano24}
Urbano, M., Duc, P.~A., Saifollahi, T., et al. \ 2024, arXiv e-prints, arXiv:2412.17672. doi:10.48550/arXiv.2412.17672
\bibitem[VanderPlas et al.(2012)]{astroML}
VanderPlas, J., Connolly, A.~J., Ivezic, Z., et al. \ 2012, cidu.conf, 47. doi:10.1109/CIDU.2012.6382200
\bibitem[Vasiliev \& Baumgardt (2021)]{Vasiliev21}
Vasiliev, E. \& Baumgardt, H. \ 2021, MNRAS, 505, 5978
\bibitem[Veljanoski et al.(2014)]{Veljanoski14}
Veljanoski, J., Mackey, A.~D., Ferguson, A.~M.~N., et al. \ 2014, MNRAS, 442, 2929
\bibitem[Veronica et al.(2025)]{Veronica25}
Veronica, A., Reiprich, T.~H., Pacaud, F., et al. \ 2025, A\&A, 694, A168
\bibitem[Walker, Fabian \& Sanders (2013)]{Walker13}
Walker, S.~A., Fabian, A.~C. \& Sanders, J.~S. \ 2013, MNRAS, 435, 3221
\end{thebibliography}
\end{document}